\documentclass[11pt]{article}
  
\usepackage{amsmath,amssymb}

\usepackage{graphics}
\usepackage{url}
\usepackage{hyperref}
\usepackage{bm}
\usepackage{slashed}
\usepackage{cite}
\usepackage{verbatim}

\usepackage{tikz}

\usepackage{color}

\newcommand{\nc}{\newcommand}  

\nc{\beq}{\begin{equation}}  
\nc{\eeq}{\end{equation}}  
\nc{\beqa}{\begin{eqnarray}}  
\nc{\eeqa}{\end{eqnarray}}  
\nc{\bit}{\begin{itemize}}  
\nc{\eit}{\end{itemize}}

\title{  
\vspace*{-2.3cm}
\begin{flushright}
\normalsize{
  }
\end{flushright}
\vspace{1.5cm}
\Large  
\textbf{Hydrogen Axion Star \\ {\large Metallic Hydrogen Bound to a QCD Axion BEC}}\vspace*{1.0cm}   
}

\author{ Yang Bai, Vernon Barger and Joshua Berger
\vspace{5mm}
\\
\normalsize\emph{Department of Physics, University of Wisconsin-Madison, Madison, WI 53706, USA}  
}
\date{}

\begin{document}  
\setcounter{page}{0}  
\maketitle  

\vspace*{1cm}  
\begin{abstract} 
As a cold dark matter candidate, the QCD axion may form Bose-Einstein condensates, called axion stars, with masses around $10^{-11}\,M_{\odot}$. In this paper, we point out that a brand new astrophysical object, a Hydrogen Axion Star (HAS), may well be formed by ordinary baryonic matter becoming gravitationally bound to an axion star. We study the properties of the HAS and find that the hydrogen cloud has a high pressure and temperature in the center and is likely in the liquid metallic hydrogen state. Because of the high particle number densities for both the axion star and the hydrogen cloud, the feeble interaction between axion and hydrogen can still generate enough internal power, around $10^{13}~\mbox{W}\times(m_a/5~\mbox{meV})^4$, to make these objects luminous point sources. High resolution ultraviolet, optical and infrared telescopes can discover HAS via black-body radiation.  
\end{abstract}  
  
\thispagestyle{empty}  
\newpage  
  
\setcounter{page}{1}  
  
\baselineskip18pt   

\vspace{-2cm}

\section{Introduction}
\label{sec:intro}
The QCD axion remains a compelling solution to both the strong CP \cite{Peccei:1977hh,Weinberg:1977ma,Wilczek:1977pj,Shifman:1979if,Kim:1979if,Zhitnitsky:1980tq,Dine:1981rt} and dark matter puzzles of the Standard Model (SM)~\cite{Turner:1989vc,Sikivie:2006ni}.  An axion that addresses both puzzles is a pseudo Nambu-Goldstone boson of a spontaneously broken Peccei-Quinn (PQ) global symmetry with a potential generated by non-perturbative QCD effects.  The minimum of the potential relaxes the effective strong CP angle to zero, resolving the strong CP puzzle.  The properties of the QCD pion sector fix a relation between the mass and decay constant of the axion.  The coupling of the axion to the SM is also fixed by these parameters, up to model-dependent coefficients that are typically $\mathcal{O}(1)$ when they do not vanish.  

Due to its weak coupling the SM, the axion is typically difficult to probe experimentally or observationally.  Astrophysically, axions can cause stellar energy losses through Primakoff production~\cite{Turner:1989vc,Raffelt:1990yz}.  The axions produced in the sun in such a way can be directly detected by helioscopes~\cite{Sikivie:1983ip} as in the CAST~\cite{Barth:2013sma} and IAXO~\cite{Armengaud:2014gea} experiments.  Axions could also convert to photons in astrophysical magnetic fields~\cite{Payez:2014xsa} or affect the dynamics of black holes through superradiance~\cite{Arvanitaki:2009fg}.  Non-astrophysical probes include photon conversion in magnetic fields, either via light shining through walls~\cite{Sikivie:1983ip}, as in the OSQAR~\cite{Ballou:2015cka} and upcoming ALPS-II~\cite{Bahre:2013ywa} experiments, or shifts in polarization~\cite{Maiani:1986md} as searched for by PVLAS~\cite{DellaValle:2014wea}.  Axions could also be detected as spin-dependent mediators of a new force~\cite{Arvanitaki:2014dfa}.

The probes mentioned above do not rely on a cosmic axion population.  If one assumes that axions make up a dominant component of cold dark matter, then additional constraints can be obtained.  Cosmic cold axion production is dominated in the simplest QCD axion models by misalignment and axion string decays, which give highly uncertain predictions for the abundance of QCD axion dark matter.  This abundance fixes another, independent relation between the mass and self-coupling of the axion and completely determines the axion potential in principle.  Given the large uncertainty on the axion abundance generated for a given set of axion parameters, there remains a range of particular interest, $10^{-5}$--$10^{-2}$~eV~\cite{Wantz:2009it}, in axion dark matter searches.  It is rather difficult to probe this range of interest because of feeble interactions of axion with SM particles, but experiments have been designed to push in that direction.

A light QCD axion is unstable and can generically decay into pairs of photons, yielding an indirect detection signal that could be seen by telescopes. Because of the suppression factor of axion mass over axion decay constant, their decay signatures are too weak for the current telescopes for an axion mass below 1~eV. In addition, if one makes some assumptions about the local density of axions on Earth, then axions can be detected directly in microwave cavity experiments like ADMX~\cite{Asztalos:2009yp}, Yale~\cite{Brubaker:2016ktl} and CULTASK~\cite{WChung,Youn:2016crv} or ABRACADABRA~\cite{Kahn:2016aff} or MADMAX~\cite{TheMADMAXWorkingGroup:2016hpc} or using magnetic resonance techniques in experiments like CASPEr~\cite{Budker:2013hfa} (see Ref.~\cite{Graham:2015ouw} for a recent review).  ADMX in particular has recently claimed the first bounds of the QCD preferred region of axion parameter space~\cite{Asztalos:2009yp}, though the application of their results to the QCD axion relies on assumptions about the abundance and local distribution of dark matter that we examine critically below.

Another potential avenue to probe axion dark matter exists that does not necessarily rely on axion coupling to the SM exists.  Assuming that the end of inflation scale is above PQ-breaking scale and before the QCD phase transition, the axion field has $\mathcal{O}(1)$ fluctuations on a length scale of $H^{-1} \lesssim 10~{\rm km}$~\cite{Guth:2014hsa}.  This large variation on small distance scales, along with the high occupancy number of axions, allows for axions to relax into a high density, compact Bose-Einstein condensate (BEC) clump conventionally known as an axion star or minicluster \cite{Hogan:1988mp,Diemand:2005vz,Zurek:2006sy}.  The exact properties of such a clump~\cite{Sikivie:2009qn,Erken:2011dz,Guth:2014hsa,Eby:2014fya,Eby:2015hsq,Braaten:2015eeu,Sikivie:2016enz,Hardy:2016mns,Hertzberg:2016tal}, as well its stability~\cite{Davidson:2016uok,Eby:2016cnq,Helfer:2016ljl,Braaten:2016dlp} has come under scrutiny recently.  Our determination is that the most likely final state for primordial QCD axion stars is a dense axion star with meter scale radius and order $10^{19}~{\rm kg}$ ($10^{-11}~M_\odot$) mass.  Such a conclusion remains to be fully tested via numerical simulations that are beyond the scope of our work.

Rather, we focus on the observational manifestations that can result from the likely structure of primordial axion stars.  Some potential signals specific to axion stars have been studied.  Axion stars are sufficiently dense that multi-axion interactions are possible, yielding boosted signals from mutli-axion processes  $n\,a \to 2\,a$ and photons from $n\,a \to 2 \,\gamma$, where $n > 2$ is the number of incoming axions in the process.  Collisions of axion stars with a dense baryonic object such as a neutron star could also generate photons and could potentially explain fast radio bursts~\cite{Iwazaki:2014wta,Raby:2016deh}.  Collapsing of axion stars could undergo maser-like axion production.  

In this article, we study signals generated by a primordial hydrogen cloud that forms around the axion star, forming a HAS.  The axion star, being a massive, dense object, seeds small scale baryonic structure formation at a very early time.  Hydrogen and helium gas gravitationally collapse on it and reach a cloud in hydrostatic quasi-equilibrium, balanced by thermal pressure.  As the universe expands, several changes happen.  The gas surrounding the cloud gets colder and collapses further, leading to growth of the cloud.  The cloud properties change from insulating gas to conducting fluid, allowing the conduction band electrons to convert axions and heat up the cloud.  The hot cloud cools back down as it expands, as the generated heat is spread wider within the cloud.  The hot cloud radiates blackbody photons from its surface, which yields the observable signal with which we are concerned.

The properties of such clouds at low redshifts below $z = 10$ are dependent on the axion star surroundings and are no longer uniform due to reionization, collapse into galaxies, and heating from astrophysical sources.  Nevertheless, so long as the cloud isn't destroyed by a collision with an unrelated molecular cloud, the power emitted should remain unchanged at roughly $10^{13}~{\rm W}$ (for $10^{-3}~\mbox{eV} \lesssim m_a \lesssim 10^{-2}~\mbox{eV}$) and be detectable by a telescope with an appropriate frequency band.  We study the potential sensitivity of such a telescope to a nearby baryon-clouded axion star, axion stars in a dwarf spheroidal satellite and diffuse emissions from baryon-clouded axion stars.  The best sensitivity would come from a nearby axion star point source, which could likely be within a few hundred astronomical units of the Earth in distance.  The signal should be much larger than the diffuse background if the axion star is located in a direction away from the galactic plane and should be readily visible to a high resolution telescope, like GAIA~\cite{GAIA}, Pan-STARRS~\cite{Pan-STARRS}, KECK~\cite{KECK} and LAMOST~\cite{LAMOST}. 

The remainder of this article is structured as follows.  In Section~\ref{sec:axion-thermal}, we review the properties of QCD axion dark matter and determine the most relevant parameter space for the axion of QCD or any new strong dynamics sector.  Then, in Section~\ref{sec:axion-stars}, we review the formation of axion stars and determine the most likely properties of primordial axion stars.  We turn to the structure of the baryonic cloud around axion stars in Section~\ref{sec:baryonic-cloud}.  We study the potential for observing HAS in Section~\ref{sec:detection}.  The consequences of an axion star inside a solar system planet are studied in Sectin~\ref{sec:axion-planets}.  Finally, in Sections~\ref{sec:discussion} and \ref{sec:conclusions}, we discuss uncertainties in our analysis and conclude.  Appendix~\ref{appendix} presents some details of the conversion of axions to heat in a metal.

\section{Axion as Cold Dark Matter}
\label{sec:axion-thermal}

In this section, we present relevant aspects of early universe axion cosmology, as well as the large scale features of axions as dark matter.  In particular, we discuss the various mechanisms for axion production in the early universe and the parameter space for axions as dark matter.  We briefly present a model that generalizes the axion parameter space beyond the QCD axion, allowing for a wider range of axion phenomenology.  Helpful reviews of axion cosmology can be found in Refs.~\cite{Turner:1989vc,Sikivie:2006ni}.

\subsection{Axion Couplings and Lifetime}
\label{sec:lifetime}
In order for the axion to be a viable cold dark matter candidate, it must first and foremost be stable.  Axion couplings to SM particles are expected to be extremely weak, though model dependent. The relevant interactions for our discussion are
\beqa
{\cal L} \supset c_\gamma\,\frac{\alpha}{4\pi\,f_a} \,a\,F_{\mu\nu}\widetilde{F}^{\mu\nu} \,-\,c_\psi\,\frac{\partial_\mu a}{f_a}\,\overline{\psi} \gamma^\mu\gamma^5 \psi \,.
\label{eq:axion-coupling}
\eeqa
Here, $\psi$ represents SM fermions. The coefficient $c_\gamma$ is related to the triangle anomaly of $U(1)_{\rm PQ}$ with two electromagnetic currents. For certain models, the anomaly could be highly suppressed such that the axion lifetime can be increased. The axion decay width is
\beqa
\Gamma(a \rightarrow \gamma \gamma) = \frac{c_\gamma^2\,\alpha^2\,m_a^3}{64\,\pi^3\,f_a^2} \,,
\eeqa
with the decay lifetime of
\beqa
\tau^0(a) &\approx& 2.5 \times 10^{24}\,\mbox{s}\,\times \,c_\gamma^{-2}\,\left(\frac{100~\mbox{eV}}{m_a} \right)^3 \, \left(\frac{f_a}{10^{10}\,\mbox{GeV}} \right)^2  \,.
\label{eq:axion-decay}
\eeqa
%

\subsection{Energy Density from the Misalignment Mechanism}
\label{sec:misalignment}
In order for an axion star to form, the PQ symmetry breaking scale must be below the inflation scale.  In parts of parameter space for which the PQ breaking happens before or during inflation, the small scale density perturbations required for primordial axion stars are washed out and the axion density is nearly uniform on small length scales of order the Hubble length.  On the other hand, if PQ breaking happens after inflation, then initial angles for the axion field are only uniform on a scale of order Hubble, with different Hubble patches having different angles sampled uniformly. As a result, the non-linear perturbations of axion field can easily generate localized clumps or miniclusters. This sets an upper bound on the decay constant of $f_a \lesssim H_I/(2\pi) \approx 10^{13}$~GeV~\cite{Ade:2015lrj}. There is also a lower bound of $f_a \gtrsim 10^8$~GeV from stellar energy-loss limits~\cite{Raffelt:1990yz,Agashe:2014kda}. So, for a light axion below ${\cal O}(1~\mbox{MeV})$, the allowed $f_a$ range is
\beqa
10^8~\mbox{GeV} \lesssim f_a \lesssim 10^{13}~\mbox{GeV} \,.
\eeqa

For the axion misalignment mechanism and in the standard cosmology (see Ref.~\cite{Fox:2004kb} for other non-standard cosmology models), the axion field equation of motion is 
\beqa
\ddot{\theta} + 3 H(T) \dot{\theta} + m_a^2(T) \theta =0\,,
\eeqa
for $\theta(x) \equiv a(x)/f_a$ and with the Hubble rate given by
\beqa
H(T) = 1.66 \, \sqrt{g_*(T)} \, \frac{T^2}{M_{\rm pl}} \,,
\eeqa
in the radiation dominated era.  Here, $g_*(T)$ is the temperature-dependent number of relativistic degrees of freedom and the Planck scale $M_{\rm pl} = 1.22 \times 10^{19}~{\rm GeV}$.
At earlier times with higher temperature, one can ignore the axion mass and have a time-independent solution for $\theta(x)$ (the other solution proportional to $1/\sqrt{t}$ decays away by later times and can be ignored). The axion mass term becomes important at a temperature of $T_1$ with $3\,H(T_1)=m_a(T_1)$, which requires us to know the temperature-dependent axion mass.

Based on a dilute instanton gas calculation~\cite{Gross:1980br,Turner:1985si,Turner:1989vc}, the axion mass increases as the temperature decreases with a power-law behavior. An updated study based on the interacting instanton liquid model shows a slightly higher mass for temperatures above the confinement scale~\cite{Wantz:2009it}. More recently, a calculation based on lattice simulations has shown that the axion mass could increase much faster than the conventional power-like behavior, although with a large uncertainty~\cite{Kitano:2015fla,Frison:2016vuc}. In this study,  we will simply follow the traditional diluted instanton gas calculation in Ref.~\cite{Turner:1985si} to estimate the oscillation temperature. 

At zero-temperature, the axion mass is
\beqa
m_a = 6\times 10^{-5}~\mbox{eV}~\times \frac{\sqrt{y_u^{-1} + y_d^{-1}} }{\sqrt{\sum_{i}^{N_f^0} y_i^{-1}}}\,
\left(\frac{v}{246~\mbox{GeV}}\right)^{1/2}\, \left(\frac{\Lambda_{\rm QCD}}{0.2~\mbox{GeV}}\right)^{3/2} \left(\frac{10^{11}~\mbox{GeV}}{f_a}\right) \,,
\eeqa
where $\Lambda_{\rm QCD}$ is the QCD strong coupling scale and $N_f^0$ is the number of light quarks with $m_i < \Lambda_{\rm QCD}$. At higher temperature, following the calculation in Ref.~\cite{Turner:1985si}, the axion mass as a function of $T$ is
\beqa
m_a^2(T) = \frac{\Lambda_{\rm QCD}^{4} }{f_a^2}\,\frac{m_1\,\cdots m_{N_f}}{\Lambda_{\rm QCD}^{N_f}}\,\left( \frac{\Lambda_{\rm QCD}}{T} \right)^{7 + N_f/3}\,I(T, \Lambda_{\rm QCD}, N_f) \,.
\eeqa
Here, $N_f$ as the number of flavor below $T$ and the function $I(T, \Lambda_{\rm QCD}, N_f)$ can be found in Eq.~(6.15) of Ref.~\cite{Gross:1980br}. After numerical integration, the axion mass can be fit by the following formula
\beqa
m_a(T) &=& \alpha \times 10^{-5}\,\mbox{eV}\,\left(\frac{v}{246~\mbox{GeV}}\right)^{\frac{N_f}{2}}\,\left(\frac{\Lambda_{\rm QCD}}{0.2~\mbox{GeV}}\right)^{2-\frac{N_f}{2}} \,\left(\frac{\Lambda_{\rm QCD}}{T}\right)^{\frac{7}{2}+\frac{N_f}{6}} \, \nonumber \\
&& \hspace{2cm} \times\left(\frac{10^{11}~\mbox{GeV}}{f_a}\right) \,\left[ 1 - \ln{\left(\frac{\Lambda_{\rm QCD}}{T}\right)} \right]^\beta \,.
\label{eq:ma-temperature}
\eeqa
The numerical values of $\alpha$ and $\beta$ for different numbers of light flavors are listed in Table~\ref{tab:a-b}. 
\begin{table}[htb!]
\renewcommand{\arraystretch}{1.8}
\begin{center}
\begin{tabular}{c|cccccc}
\hline\hline
  $N_f$ & 1 & 2 & 3 & 4 & 5 & 6  \\  \hline
   $\alpha$   & 1.7 & 0.21 & 0.15 & 0.25 & 0.71 & 5.9  \\ \hline
   $\beta$   & 0.84  & 1.02 & 1.22 & 1.46  & 1.74  & 2.07 
 \\ \hline  
 \hline
\end{tabular}
\end{center}
\caption{The numerical values of $\alpha$ and $\beta$ for $m_a(T)$ in Eq.~(\ref{eq:ma-temperature}). 
\label{tab:a-b}}
\end{table}%

The misalignment mechanism works as follows.  Given the initial constant value of the axion field in a given Hubble patch and the negligible axion potential well before the QCD phase transition, the energy density of the axion field is initially negligible.  At a temperature $T_1$, the axion mass becomes larger than Hubble and the axion field begins to roll toward its minimum, while the energy density in the axion field grows for this time-dependent Hamiltonian.  Around the QCD phase transition temperature, the energy density in the axion field per comoving volume freezes out and the axion field has relaxed to its minimum.  The resulting energy density of axions at that time is 
\beqa
\rho = \frac{1}{2}\,m_a\, m_a(T_1) \,f_a^2\, \langle\theta_i^2 f(\theta_i) \rangle\,,
\eeqa
where $f(\theta_i)$ takes into account of the anharmonic terms in the axion potential and other $\mathcal{O}(1)$ effects.  
The energy density scales with the entropy density to its present day value of~\cite{Sikivie:2006ni} 
\beqa
\rho^{\rm mis-align} = \frac{m_a \, m_a(T_1)\,s(T_0) }{2\,s(T_1)}\,f^2_a \langle \theta_i^2 f(\theta_i) \rangle \,, 
\eeqa
where the entropy density is $s(T) = 2\pi^2 g_{*S}(T) T^3/45$ and the Hubble patch averaged initial angle dependence is given by $\langle \theta_i^2 f(\theta_i) \rangle \approx 8.77$~\cite{Visinelli:2009zm}. The dark matter energy fraction parameter $\Omega_a = \rho_a / \rho_{\rm crit}$ with $\rho_{\rm crit} = 3 H_0^2 M^2_{\rm pl} / 8\pi$, to be matched to the experimental value of $\Omega_{\rm CDM} h^2 = 0.1199\pm 0.0022$ and $h = 0.678\pm 0.009$~\cite{Ade:2015xua} if the axion is the only component of dark matter.

\subsection{Energy Density from Axion String Decay}
\label{sec:string-decay}
\begin{figure}[th!b]
  \centering
\includegraphics[width=0.65\textwidth]{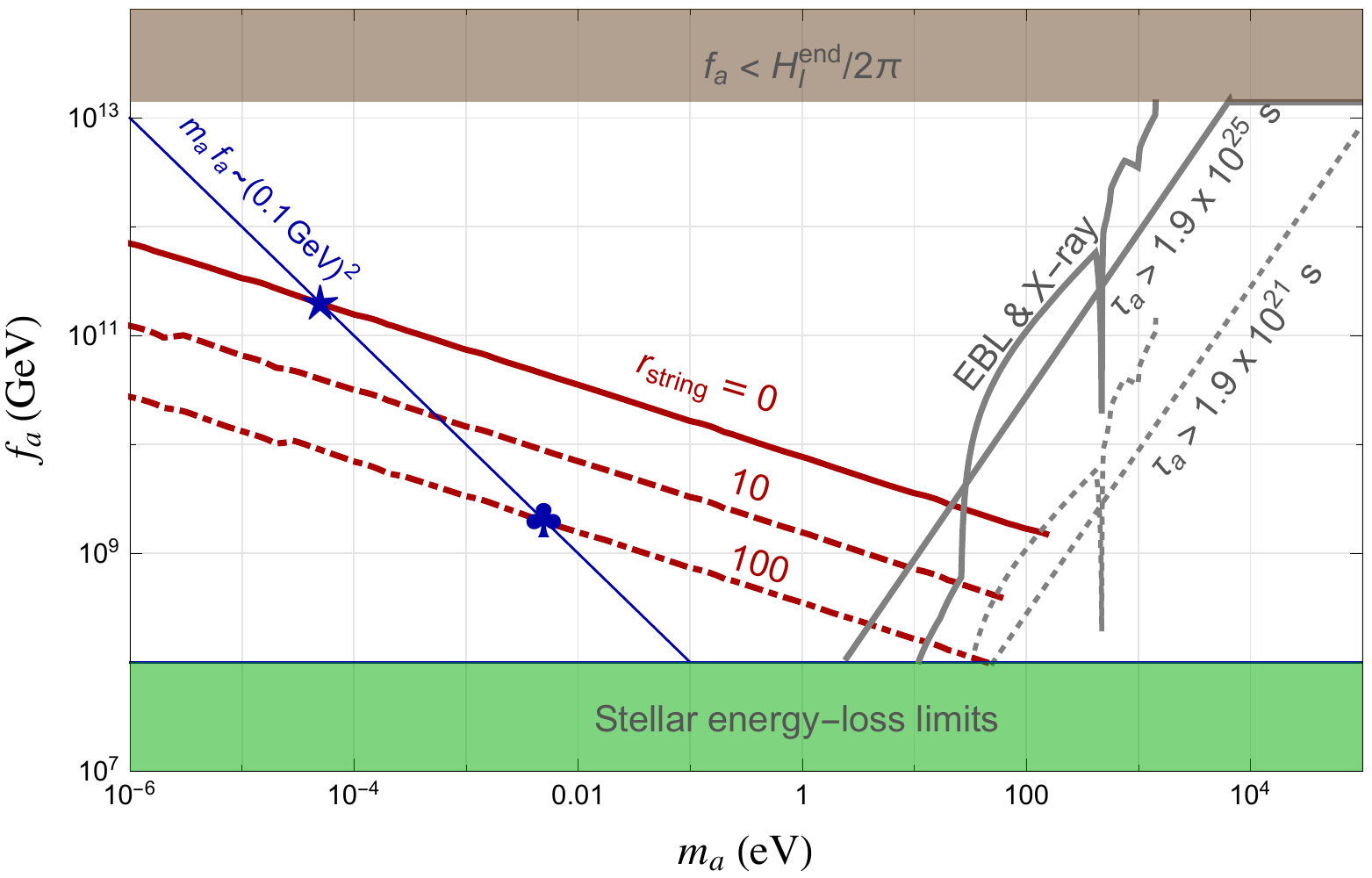}
  \caption{The parameter space in which QCD axions account for all cold dark matter in the universe for different values of $r_{\rm string}$. The decay constant $f_a$ is bounded from above by the Hubble scale at the end of inflation~\cite{Ade:2015lrj} and from below by the stellar energy loss~\cite{Agashe:2014kda}. The parameter space on the right side of EBL \& X-ray curves are excluded for $c_\gamma=1$, where $c_\gamma$ is defined in Eq.~\eqref{eq:axion-coupling}. The region on the right side of the gray and straight line labeled by $\tau_a > 1.9 \times 10^{25}$~s is excluded by reionization effects on CMB observables. The gray and dashed lines are for $c_\gamma =0.01$ and a smaller coupling of axion to two photons. The two benchmark points have: $(m_a, f_a)=(5\times 10^{-5}~\mbox{eV}, 2\times10^{11}~\mbox{GeV})$ (star) and $(5\times 10^{-3}~\mbox{eV}, 2\times10^{9}~\mbox{GeV})$ (cloud). }
  \label{fig:ma-fa}
\end{figure}

As alluded to in the previous section, the value of the axion field is only constant on roughly the Hubble scale.  At the PQ phase transition, topological defects may form on that length scale.  We focus on the contributions to the axion abundance due to axion string defects.\footnote{Note that, in some models, the ones considered in this work in particular, the model as written thus far would also include domain walls even after the QCD-induced potential is included.  These arise due to residual $Z_N$ symmetries of the potential.  The domain walls are generally inconsistent with observed cosmology and need to be destabilized.  This can be achieved, for example, by introducing a small $Z_N$ breaking potential term~\cite{Abel:1995wk,Chung:2010cd}.}
Axion strings can be produced in early universe by the Kibble mechanism~\cite{Kibble:1976sj,Hindmarsh:1994re} and contribute non-negligibly to the energy density for the $U(1)_{\rm PQ}$ breaking after inflation. Two types of decaying strings produce axions: loop strings and long strings. Detailed axion production results rely on the typical loop size and the back-reaction rate~\cite{Sikivie:2006ni,Wantz:2009it}. Depending the axion energy spectrum from string radiating axions, the ratio of axion production from string decays to that from mis-alignment can vary up to two orders of magnitude. The abundance of axions generated nevertheless has the same axion mass and decay constant parametric dependence~\cite{Wantz:2009it}, so that we can simply parametrize the theoretical uncertainty in string decay.  We therefore define the axion parameter-independent ratio
\beqa
r_{\rm string} = \frac{\rho^{\rm string-decay} }{\rho^{\rm mis-align}} \,,
\eeqa
and take the summation of the two contributions as the total axion energy density. In Fig.~\ref{fig:ma-fa}, we show the parameters $m_a$ and $f_a$ required for axions to account for all the cold dark matter in the universe. Generically, if the contribution from string decays is increased, a lower value of $f_a$ is needed for a fixed value of $m_a$. We also show the SM QCD axion relation of $m_a f_a \sim (0.1~\mbox{GeV})^2$ in the blue straight  line of Fig.~\ref{fig:ma-fa}. 

Also shown in Fig.~\ref{fig:ma-fa} are the constraints on the decay constant from stellar energy-loss limits~\cite{Agashe:2014kda} and requiring it below the Hubble scale at the end of inflation~\cite{Ade:2015lrj}. For a heavier axion mass between 1-100 eV, even though the axion lifetime is greater than the age of our Universe, the axion still decays fast enough to have non-trivial contributions to the extragalactic background light (EBL) and X-ray. Using the summarized results in Refs.~\cite{Overduin:2004sz,Arias:2012az}, we show the constraints on our axion model parameters in gray and curved lines. Furthermore, the decays of axions in the early Universe add additional contribution to the radiation energy and reionization effects. The axion particles decay 100\% into electromagnetic components. The cosmic microwave background (CMB) observations impose a non-trivial bound on its decay lifetime to be $\tau^0_a > 1.9\times 10^{25}$~s. We translate this constraint as gray and straight lines for two different choices of $c_\gamma = 1$ and $0.01$ in a later Eq.~(\ref{eq:axion-decay}). All together, the axion can account for all of cold dark matter and satisfy experimental constraints if its mass falls in the range ${\cal O}(10^{-5}) - {\cal O}(100)$~eV.  Generally, we consider models in which SM QCD generates the dominant potential for the axion.  It may be phenomenologically interesting, however, to consider a more general parameter space with more massive axions.  In order to move away from the QCD axion line in the plot, additional contributions to the axion potential are required, which we briefly discuss below.

\subsection{Heavy Axion Model}
\label{sec:axion-model}
In a typical axion model, the $U(1)_{\rm PQ}$ symmetry is spontaneously broken such that the corresponding pseudo-Nambu-Goldstone boson $\phi$ has the shift symmetry of $\phi \rightarrow \phi + \epsilon\,f_a$ under the infinitesimal $U(1)_{\rm PQ}$ transformation. The $U(1)_{\rm PQ}$ is also anomalous under the SM QCD interaction with a non-zero triangular anomaly, $A^c_{\rm PQ} \equiv \mbox{Tr}(t^a t^b Q_{\rm PQ})$. The QCD instanton effects generate an effective potential for the axion field $V(\phi)\sim m_\pi^2 f_\pi^2 \cos(\phi/f_a - \theta)$. Minimizing this potential, one has $\langle \phi \rangle = f_a\,\theta$, which makes the physical strong-CP parameter $\overline{\theta}_{\rm eff} = \overline{\theta} - \phi/f_a$ zero and solve the strong-CP problem. 

For the ``invisible axion'' models with $f_a \gg f_\pi$, the mixings of axion with $\eta^\prime$, $\eta$ and $\pi^0$ are suppressed by $f_\pi/f_a$ and are small. The axion mass is model-dependent and related to the quark masses as~\cite{Kaplan:1985dv}
\beqa
 m_a \, f_a = \frac{\sqrt{m_u/m_d}}{1 + m_u/m_d}\,m_\pi\,f_\pi\,A^c_{\rm PQ} \,.
\eeqa
So, the axion mass $m_a$ and its decay constant $f_a$ are not independent parameters and are related by SM QCD scales, up to order-one model parameters. 

Beyond the minimum SM QCD axion models, additional models aim to increase the axion mass. For instance, in Ref.~\cite{Rubakov:1997vp}, a hidden QCD sector is considered to have additional instanton contributions to the axion mass (see also Ref.~\cite{Hook:2014cda} for a recent study). The effective potential for the pseudo-Nambu-Goldstone boson $\phi$ becomes
\beqa
V(\phi) = c\,m_\pi^2 \,f_\pi^2\,\cos{\left(\frac{\phi}{f_a} - \theta\right)}  + c^\prime\,m_{\pi'}^2 \,f_{\pi'}^2\,\cos{\left(\frac{\phi}{f_a} - \theta'\right)} \,,
\eeqa
where $c$ and $c'$ are model-dependent number of order unity and $\theta'$ is the $\theta$-angle of the hidden QCD sector. To solve the strong-CP problem, one also need to require $\theta = \theta'$. This can be achieved by assuming a ${\cal Z}_2$ symmetry in the UV theory such that the $\theta$-angles in both sectors are identical. After imposing the vacuum expectation value (VEV) of $\phi$ field, we have the potential for the axion field as
\beqa
V(a) = \left( c\,m_\pi^2 \,f_\pi^2\,+\,c^\prime\,m_{\pi'}^2 \,f_{\pi'}^2\right) \,\cos{\left(\frac{a}{f_a}\right)} \,.
\label{eq:axion-potential}
\eeqa
For the parameter space with $c^\prime\,m_{\pi'}^2 \,f_{\pi'}^2 > c\,m_\pi^2 \,f_\pi^2$, the axion mass is dominated by the hidden sector contributions with $m_a\,f_a \approx \sqrt{c'}\,m_{\pi'} f_{\pi'}$. Absent a picture of hidden QCD dynamics, we should treat $m_a$ and $f_a$ as independent parameters. As seen above, requiring axions to account for the entire dark matter density fixes one relation between $m_a$ and $f_a$, up to theoretical uncertainties.  

To calculate the axion relic energy density, it is also important to know the time-dependent mass, which cares about the hidden quark masses in the hidden QCD sector. One nature choice based on the ${\cal Z}_2$ symmetry is to have the hidden QCD with $N_c'=3$ with the total number of flavor $N_f'=6$. The Yukawa couplings in the hidden sector are identical to the visible sector: $y_f^i = y_{f'}^i$. On the other hand, the QCD confinement and electroweak scales in both sectors could be different. Generically, we will assume $v' > v$ and $\Lambda'_{\rm QCD} > \Lambda_{\rm QCD}$ to have the relation $c^\prime\,m_{\pi'}^2 \,f_{\pi'}^2 > c\,m_\pi^2 \,f_\pi^2$ in Eq.~(\ref{eq:axion-potential}). 

\section{Axion Stars}
\label{sec:axion-stars}
For the two main non-thermal axion production mechanisms, the axions are extremely cold after the QCD phase transition~\footnote{Throughout this description, we describe phenomena in terms of QCD for simplicity, though by QCD we may also mean a QCD-like QCD$'$.} and may form a Bose-Einstein condensate. Furthermore, because of the large inhomogeneity of axion energy density at or below the oscillation temperature $T_1$, the group of axions that are causally connected within the Hubble volume at $T_1$ can be gravitationally bound to form axion stars, sometimes also called axion miniclusters~\cite{Hogan:1988mp,Kolb:1993zz,Kolb:1995bu}. Numerical simulations are in general needed to answer questions like the fraction of axions belong an axion star and the mass and size of the axion stars. Some initial attempts to answer those questions can be found in Refs.~\cite{Kolb:1993zz,Khlebnikov:1999qy}. In this paper, we will mainly concentrate on the potential phenomenological consequence of axion stars and will only have estimations of some axion star properties at the order-of-magnitude level. The general picture of axion stars arises from the following timeline:
\begin{itemize}
\item Reheating after inflation places the axion field at a
  temperature above the PQ phase transition that occurs at $T_{\rm PQ}
  \sim f_{\rm PQ}$.  There are small inhomogeneities due to quantum fluctuations in the (now decayed) inflaton.
\item At $T_{\rm PQ}$, the axion field ({\it the radial field}) rolls down to it's minimum in a
  random $U(1)_{\rm PQ}$ phase direction that is homogeneous on a scale of roughly $H(T_{\rm PQ})^{-1}$.  Excitations along the flat direction at the bottom of the Mexican-hat potential are the axions.  At this stage, they can be treated as a massless classical scalar field in an expanding background.  Differences in the vacuum phase throughout space lead to topological defects, namely strings.
\item As the Universe cools below the phase transition, the axion strings can split off loops that decay and can also radiate axions. Production is dominated at low temperatures and so low energies. The thermal abundance of axions is subdominant.
\item Around the oscillation temperature $T_1$ defined by $3 H(T_1) = m_a(T_1)$, the potential for
  the axions becomes sufficiently large that they roll down toward their minimum, damped by the Hubble friction.  After a few oscillations around the minimum, the axion field relaxes to its minimum value.  This
  rolling causes copious axion production.  In addition, as the mass becomes large, the string network annihilates and radiates to now massive axions and disappears.
\item At the time of the QCD phase transition, where $T_{\rm QCD}$ is
  slightly below $T_1$, the axion string network has disappeared, the
  axion mass is frozen at its zero-temperature value, and the
  abundance of axions in a comoving volume is fixed.  Since axion string radiation is dominated by low frequencies
  and the number of axions radiated per comoving volume during axion rolling is fixed, the axion population is deeply non-relativistic.
  The $\mathcal{O}(1)$ differences in the vacuum angle between
  different patches around this time lead to $\mathcal{O}(1)$
  differences in the number density of axions in those patches.  This
  inhomogeneity can lead to gravitational collapse on scales below about $H(T_1)^{-1}$ and the formation of ``{\it axion stars}''.
\end{itemize}

The axion quartic self-interaction, and later the gravitational self-interaction, are sufficient to cause the axion to locally relax into a BEC~\cite{Guth:2014hsa}.  The resulting BEC axion star contains at most the axions within roughly a Hubble volume at the QCD phase transition and is coherent at most on a Hubble length at the QCD phase transition, though we will show that the typical size of the axion star is far below that.  The ground state of the system is set by the gravitational and self-interaction potential of the clump.  Almost all the axions in the collapsing region end up in the ground state, as it is a BEC state. We now estimate the properties of the resulting axion clump. After the number of axions per comoving volume has frozen, the number density of axions is fixed by requiring that they account for the total present day energy density of dark matter $\Omega_{\rm CDM}h^2 \approx 0.12$.  For this to be the case, the number density must satisfy
\beq
n_a(T) = \Omega_{\rm CDM} \frac{\rho_{\rm crit}}{m_a} \frac{s(T)}{s(T_0)} \,,
\eeq
for the temperature from around $T_1$ to the present day temperature $T_0=2.7$~K, since the axion number per comoving volume is approximately conserved. The number of axions in a Hubble volume at the time of the QCD phase transition is thus
\beqa
\mathcal{N}_a\, =\, \frac{n_a(T_{1})}{H^3(T_1)} \, &=& \frac{\Omega_{\rm CDM}\,\rho_{\rm crit}}{1.66^3\,g_{*S}(T_0)\,g^{1/2}_*(T_{\rm 1})\,m_a}\,\frac{M_{\rm pl}^3}{T_0^3\,T_1^3} \nonumber \\
  &\approx&  4 \times 10^{57}\, \times \left[\frac{10}{g_*(T_1)}\right]^{1/2}\, \left( \frac{5\times 10^{-3}\,\mbox{eV}}{m_a} \right)^{5/2} 
     \left( \frac{2\times 10^{9}\,\mbox{GeV}}{f_a} \right)^{3/2}
\,.
\eeqa
Here, we have used $g_{*S}(T_0) = 3.91$, $g_{*S}(T_1) \approx g_{*}(T_1)$ and $T_1 \approx 1~{\rm GeV}$ from solving the relation $m_a(T_1) = 3 H(T_1)$ in the dilute instanton gas approximation.  This number of axions is roughly the amount available for gravitational collapse. We  assume that the final number of axions in an axion star is around $\mathcal{N}_a$, even through additional numerical simulations are required to know the exact relation.  That is, we assume that the fraction of axions collapsing into axion stars is 100\%.\footnote{Some simulations  have shown that under some conditions an $\mathcal{O}(1)$ fraction of axions do in fact collapse into ``dew'' drops of BEC~\cite{Khlebnikov:1999qy}.} In addition, the above estimate assumes an average Hubble patch.  Clearly, some patches will be more or less dense than others; without such an inhomogeneity, collapse is not even possible.  On the other hand, the variations compared to the mean are at most $\mathcal{O}(1)$ since the height of the axion potential is finite. 
\begin{figure}[th!b]
  \centering
\includegraphics[width=0.42\textwidth]{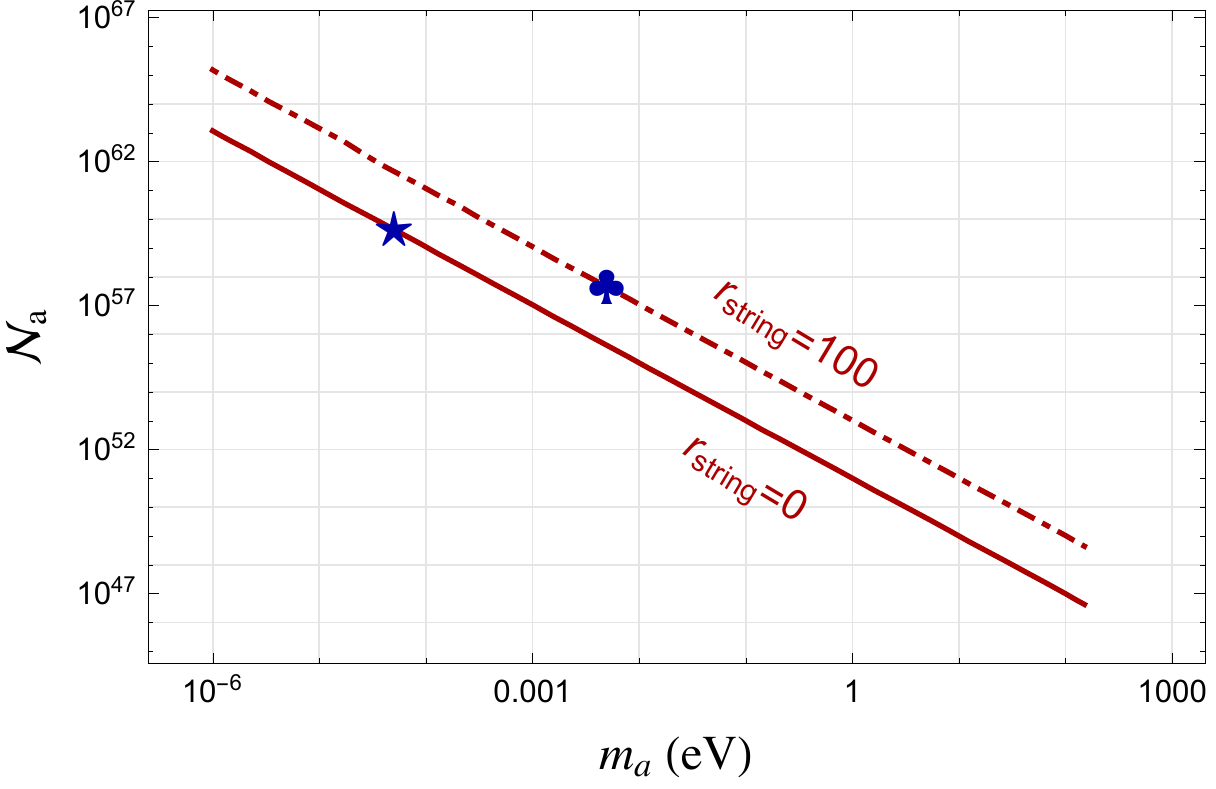} \hspace{5mm}
\includegraphics[width=0.52\textwidth]{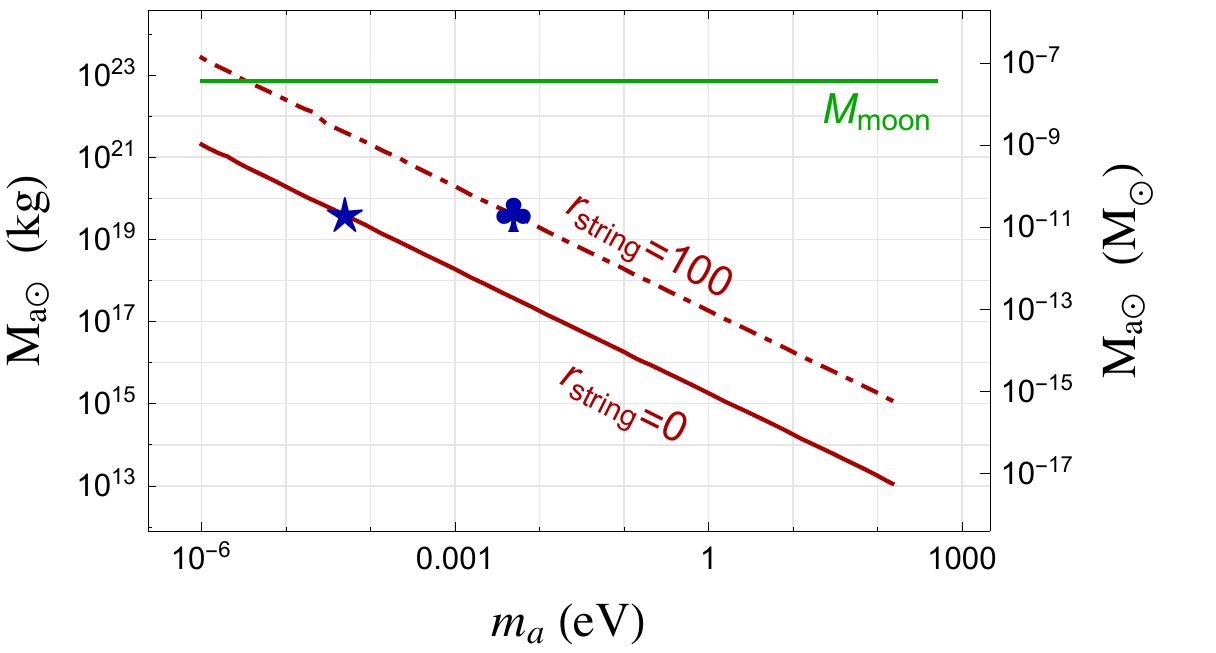}
  \caption{Left panel: the total number of axions in a typical axion star for different axion masses. The blue star and cloud denote the SM QCD axion model points for $r_{\rm string} =0$ and $r_{\rm string} =100$. Right panel: the typical axion star total mass. The axion star mass can vary from $10^{-10}\,M_{\odot}$ to $10^{-17}\,M_{\odot}$ for $m_a$ from $5\times 10^{-5}$~eV to 100~eV. For the two SM QCD benchmark points, $M_{a\odot} \approx 1.8\times10^{-11}M_{\odot}$.  }
  \label{fig:star-number-mass}
\end{figure}
\begin{figure}[th!b]
  \centering
  \includegraphics[width=0.45\textwidth]{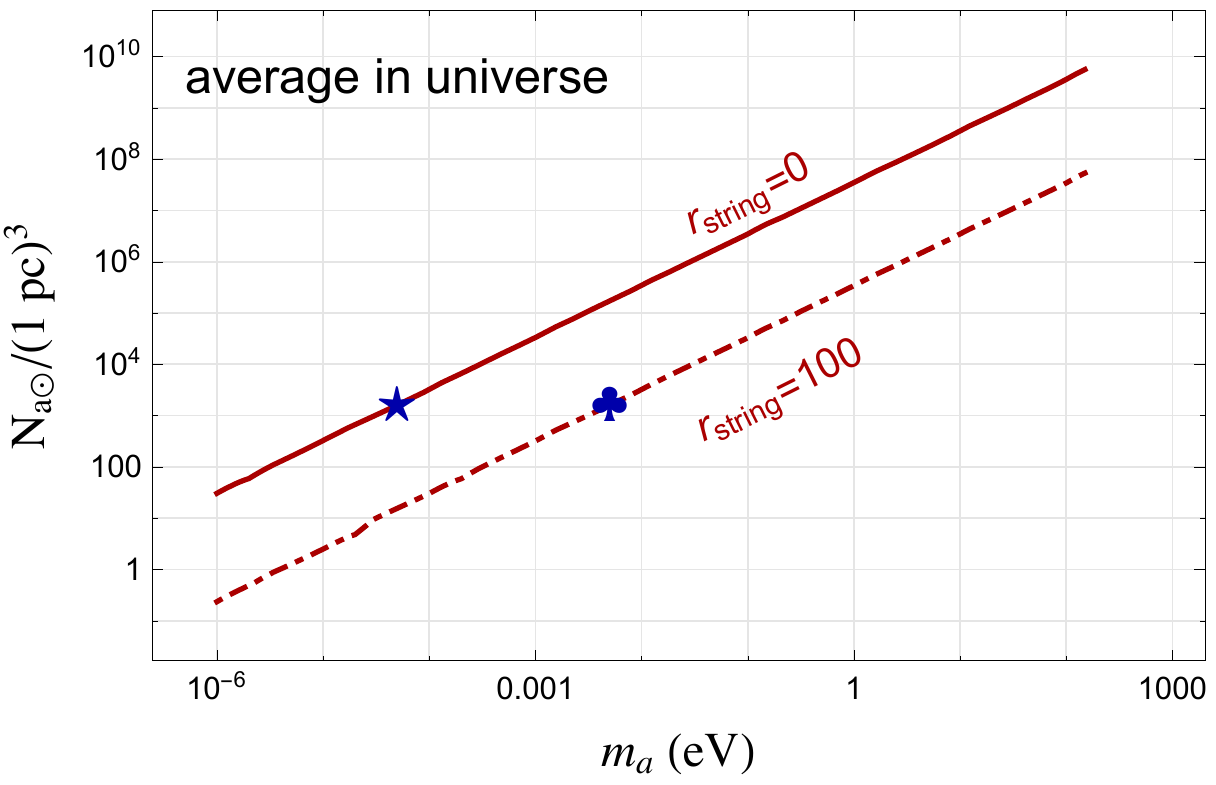} \hspace{6mm}
\includegraphics[width=0.45\textwidth]{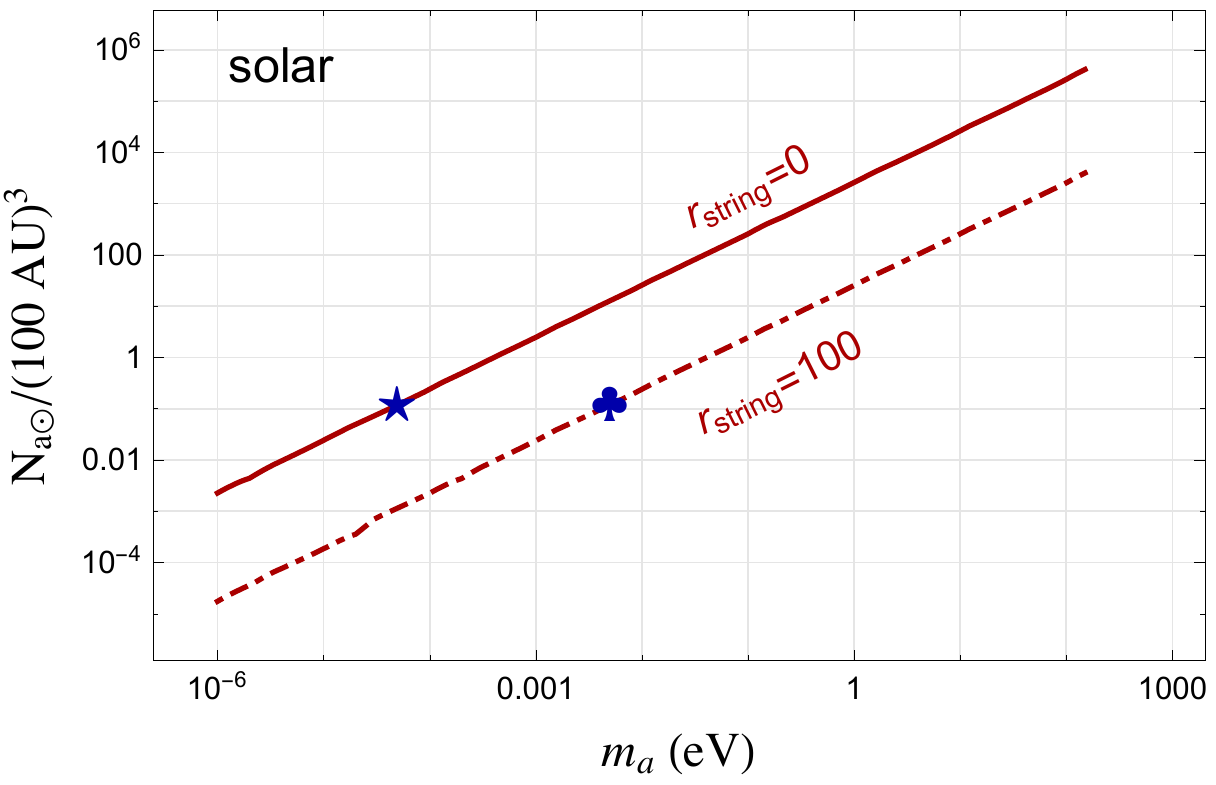}
  \caption{Left panel: the population of axion stars with an average dark matter energy density around 1.2~GeV/m$^3$ in the Universe. Right panel: the same as the left but for the local solar system dark matter energy density of $0.4$~GeV/cm$^3$.}
  \label{fig:star-size-population}
\end{figure}
The axion-star total mass is simply
\beqa
 M_{a\odot} \approx m_a \, \mathcal{N}_a = 1.8\times 10^{-11}\,M_{\odot}\, \times \left[\frac{10}{g_*(T_1)}\right]^{1/2}\, \left( \frac{5\times 10^{-3}\,\mbox{eV}}{m_a} \right)^{3/2} 
     \left( \frac{2\times 10^{9}\,\mbox{GeV}}{f_a} \right)^{3/2}\,,
 \eeqa
 which scales as $(m_a f_a)^{-3/2} \sim \Lambda^{-3}_{\rm QCD}$. 
 In Fig.~\ref{fig:star-number-mass} we show the distributions of $\mathcal{N}_a$ and $M_{a\odot}$ for different axion masses after requiring the axions to account for all dark matter. In Fig.~\ref{fig:star-size-population}, we show the population of axion stars using both the averaged dark matter energy density in the Universe and in the local solar system.

Relaxation can begin once the axion acquires a significant mass.  The relaxation time to reach a BEC state can be estimated by the time scale for evolution of the occupancy number in mode $k$.  As shown in \cite{Guth:2014hsa}, this evolution time is given by
\beqa
\tau_{k,\rm relaxation} \sim {\rm min}\,\left[\,\frac{4\,f_a^2}{n_{\rm ave}}\,,\, \frac{k^2}{8 \,\pi \, G_N \, m_a^2 \, n_{\rm ave}}\,\right] \,,
\eeqa
where $n_{\rm ave}$ is the average axion number density and $G_N=6.7\times 10^{-39}$~GeV$^{-2}$ is Newton's gravitational constant.  The number density is at least the average number density of axions in the universe, $n_a(T)$, which at $T_1$ is given by $n_a(T_1) \sim 5 \times 10^{49}~{\rm m}^{-3}$ for an axion mass of $m_a = 5 \times 10^{-3}~{\rm eV}$.  At a reference wavenumber determined by the Hubble scale at the $T_1$, relaxation is dominated by self-interactions; the relaxation time determined by gravitational interactions is more than three orders of magnitude larger.  We then find that the relaxation time is given by
\beqa
\tau_{\rm relaxation} \lesssim 2.8 \times 10^{-8}~{\rm sec} \times\frac{5 \times 10^{-3}~{\rm eV}}{m_a} \,,
\eeqa
for a QCD axion.  Note that
\beqa
\tau_{\rm relaxation} H \lesssim 0.02 \,\times\, \left[\frac{10}{g_*(T_1)}\right]^{1/2}\, \left(\frac{1~{\rm GeV}}{T}\right)\,,
\eeqa
so the axion star relaxation is rapid around the QCD phase transition.  Even though the relaxation due to self-interactions becomes slower, the gravitational relaxation of modes on the Hubble scale becomes more important and helps ensure rapid relaxation at all times. 

The gravitational collapse is halted in one of two possible ways.  If the axion star is not too massive, then quantum pressure is sufficient.  A state where the gravitational force is balanced by quantum pressure can be achieved for axion star mass $M_{a\odot} \lesssim 10.15 \,f_a / (m_a G_N^{1/2})$~\cite{Breit:1983nr,Chavanis:2011zm,Braaten:2015eeu}.  Numerically, we find $M_{a\odot} \lesssim 4 \times 10^{-13}\,M_\odot$ for $r_{\rm string} = 0$ and $m_a = 5 \times 10^{-5}~{\rm eV}$.  The upper bound decreases for $r_{\rm string} > 0$, so it is clear that a quantum pressure supported axion star is not possible if all the axions collapse into axion stars.  

\begin{figure}[th!b]
\centering
\includegraphics[width=0.48\textwidth]{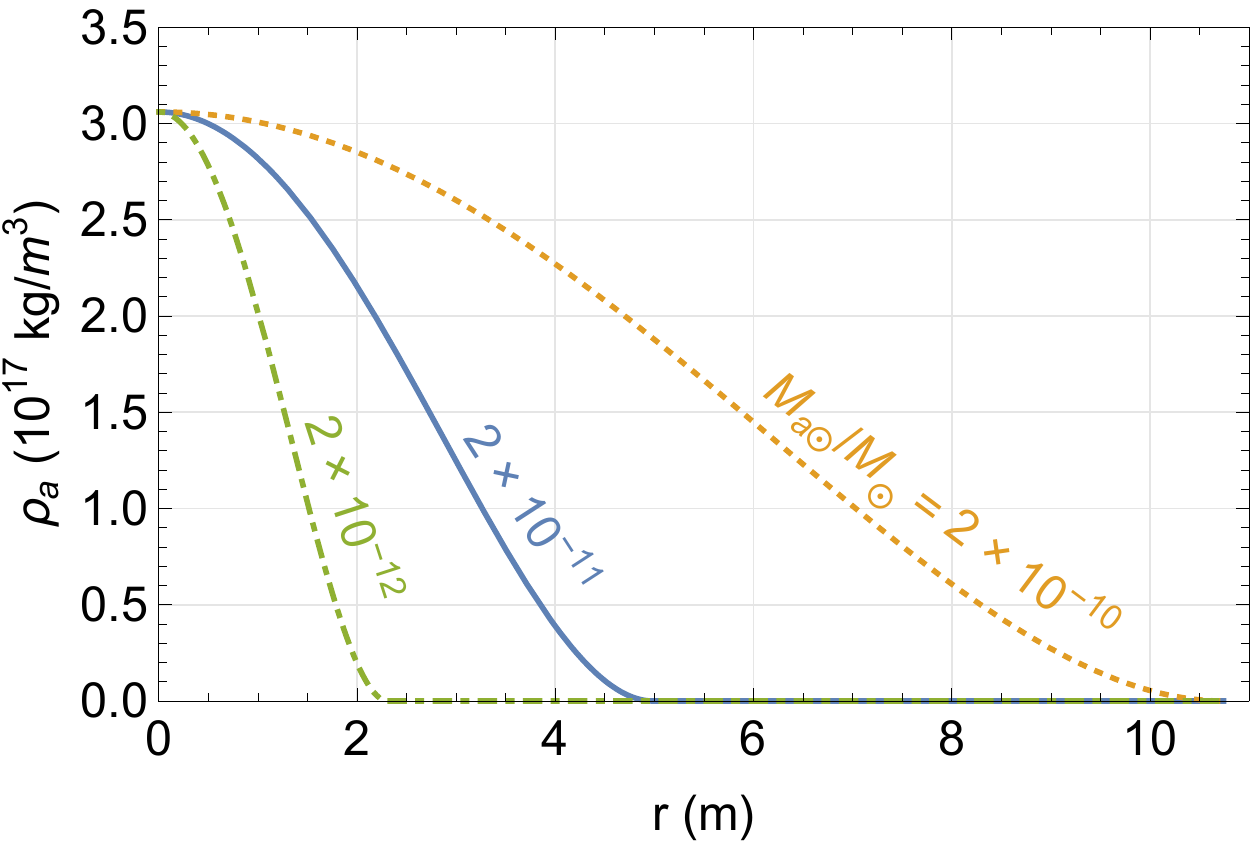}
\caption{Axion density profile for a typical axion star with different axion-star masses.\label{fig:as-proflle}}
\end{figure}

The states we consider thus fall into the ``dense'' regime studied in Ref.~\cite{Braaten:2015eeu}.  In this regime, the gravitational interactions are balanced not by quantum pressure, but rather by the self-interaction potential.  The mass of the axion star, even in the dense regime, is dominated by the mass of the axions in the star, such that $M_{a\odot} \approx m_a\, \mathcal{N}_a$.  The radius and mass of the axion star are roughly related by \cite{Braaten:2015eeu}
\beqa
\label{eq:astar-radius}
R_{a\odot} \approx 2.1 \times 10^{-5}\,R_\odot\, \left(\frac{M_{a\odot}}{M_\odot}\right)^{0.305} \approx 
6\,\mbox{m}\,\times\,\left(\frac{M_{a\odot}}{10^{-11}\,M_\odot}\right)^{0.305} 
\,,
\eeqa
which is independent of $m_a$ for fixed $m_a \,f_a$. Following Ref.~\cite{Braaten:2015eeu}, we solve for the wavefunction numerically under the Thomas-Fermi approximation, which is a good approximation in the region of interest.  The total energy is given by 
\beqa
E = \int d^3x \left[\frac{|\nabla \psi|^2}{2\,m_a} + m_a\,\Phi\,|\psi|^2 + V_{\rm eff}(|\psi|^2) \right]\,,
\eeqa
where $\Phi$ is the gravitational potential and $V_{\rm eff}$ is the effective self-interaction potential for the axion in the non-relativistic limit.  The potential can be derived in the non-relativistic limit by expanding the axion field
$a$ as $a({\bf x}, t) = [\psi({\bf x}) e^{-i m_a t} + \psi^*({\bf x}) e^{i m_a t}]/\sqrt{2m_a}$ and collecting the terms that are time independent, {\it i.e.}~slowly
varying.  The resulting potential, as in
\cite{Eby:2016cnq,Braaten:2016dlp}, is given by
\beqa
V_{\rm eff}(|\psi|^2) = m_a^2 f_a^2 \left[1 + \frac{|\psi|^2}{2\, m_a\, f_a^2} - J_0\left(\frac{\sqrt{2}\, |\psi|}{\sqrt{m_a} \,f_a}\right)\right],
\eeqa
where $J_0(x)$ is a Bessel function. The equations of motion are the Schr{\" o}edinger equation combined with the Poisson equation for gravitational potential
\beqa
\mu\,\psi = - \frac{\nabla^2 \psi}{2\,m_a}\, + \, m_a\,\Phi\,\psi\, +\, \left[V_{\rm eff}'(|\psi|^2)-m_a\right]\,\psi \,, \qquad
\nabla^2 \Phi = 4\pi G_N\, m_a\, |\psi|^2 \,.
\eeqa
Here, the space-independent chemical potential, $\mu$, does not enter our approximation for the wavefunction.
In the Thomas-Fermi approximation, the kinetic energy term is dropped, so that a single differential equation for $\psi$ is obtained, namely
\beqa
\nabla^2\,\left[V_{\rm eff}'(|\psi|^2)\right] = - 4 \pi G_N\,m_a^2\, |\psi|^2 \,.
\eeqa
The wavefunction depends only on the radial coordinate $r$ under the assumptions we have made. The boundary conditions are taken to be $|\psi(0)| = \psi_0 = n_0^{1/2}$ with $n_0$ as the central number density and $\psi^\prime(0) = 0$. As clarified in Ref.~\cite{Braaten:2016dlp}, if $V^{\prime\prime}_{\rm eff}(\psi_0^2) < 0$, then the density increases away from the origin and there is no solution that goes to 0 at infinity, that is no physical solution exists, so there are only viable solutions for $V^{\prime\prime}_{\rm eff} > 0$. The second derivative of the potential is given by a regularized hypergeometric function
\beqa
V_{\rm eff}^{\prime\prime}(\hat{n}) = - \frac{m_a^2\,f_a^2}{16}\,{}_0\widetilde{F}_1(3,- \hat{n}/4) \,,
\eeqa
with $\hat{n} \equiv 2 |\psi|^2/m_a f_a^2$; this has an infinite number of zeroes, leading to an infinite number of solution branches (one for each zero).  If the axion star collapse is sufficiently slow, it should collapse down to the least dense stable state, which is the described by a radius in Eq.~\eqref{eq:astar-radius}.  If the collapse process is more violent, the axion star could shed mass or skip down to an even more dense state.  A full simulation of the dynamics of axion star collapse is required to determine the ultimate axion star state, but for the purposes of this work we assume that the collapse is sufficiently slow that the least dense of the dense axion star states is reached.  The axion star profile depends only on the axion star total mass, which we fix as above for the QCD axion and observed dark matter abundance to $M_{a\odot} = 3.6 \times 10^{19}~{\rm kg}= 1.8\times10^{-11}M_{\odot}$.  The axion star density profiles for representative values of $M_{a\odot}$ are shown in Fig.~\ref{fig:as-proflle}.
%

\section{Primordial Baryonic Cloud around Axion Stars}
\label{sec:baryonic-cloud}
As discussed above, the axion stars are formed at a very early time in our Universe, near the QCD phase transition time. The existence of axion stars can provide strong gravitational potentials for ordinary baryons. As a result, some of the gas in the early universe can collapse into the gravitational potentials provided by the axion stars, forming a HAS. The baryonic cloud grows as the outside gas cools. Before recombination, this gas is primarily ionized protons and electrons, while after recombination it is dominantly hydrogen gas. Around 25\% is helium, but we neglect the contribution of helium in the calculations below.  From this point on, we restore factors of Planck's constant $h$ (as well as its reduced form $\hbar$), the speed of light $c$, the Boltzmann constant $k_B$ and the permittivity of free space $\epsilon_0$ for clarity.

We are primarily interested in the final state of the baryons in the axion star and not in the rapid dynamics by which they form.  By the time of recombination, all dynamical processes occur on a time scale much shorter than a Hubble time, so the equilibrium approximation is justified. For example, the free-fall time is given by
\beqa
t_{\rm ff} \lesssim \left(\frac{R_{a\odot}^3}{G_N\, M_{a\odot,\rm enc}}\right)^{1/2} \sim 2 \times 10^{-4}~{\rm s} \,,
\eeqa
where $M_{a\odot,{\rm enc}}$ is the enclosed mass of axions, which can be determined from its density profile in Fig.~\ref{fig:as-proflle}. We therefore solve for the equilibrium state of the baryons.  We are also interested in this state after recombination and so consider a fluid of hydrogen. It turns out that the hydrogen will become sufficiently dense as to reach a state of high-conductivity metallic hydrogen. In the presence of this metallic hydrogen, the axions will convert off conduction band electrons, generating a non-trivial amount of heat. We therefore must solve for thermal, hydrostatic, and heat flow equilibrium.

Thermal equilibrium enforces that there is a locally well-defined temperature $T$, so that we can make use thermodynamic properties of hydrogen such as its equation of state for pressure $P(\rho,T)$, where $\rho$ is the local mass density.  Assuming that there is no net flow in the hydrogen fluid, the condition for hydrostatic equilibrium is then
\beqa
\label{eq:hydro-eq}
\frac{\nabla P}{\rho} = - \nabla \Phi \,,
\eeqa
where $\Phi$ is the gravitational potential.  The self-gravitation of the hydrogen gas will turn out to be negligible, so that the gravitational force per unit mass is given by
\beqa
- \nabla \Phi = - \frac{G_N \,M_{a\odot,{\rm enc}}(r)}{r^2} \,\hat{r} \,.
\eeqa
We assume an isotropic distribution for axions inside the axion star as above.  At the high pressures reached for baryons in the core of the axion star, there is a series of phase transitions during which the hydrogen gas goes from an atomic gas to a molecular gas (second order), then a gas to a liquid (second order) and then from a molecular liquid to an atomic metallic liquid (generally first order)~\cite{RevModPhys.84.1607}.  For a relatively cold axion star, it is possible that some of the ``cloud'' will actually be solid hydrogen.  Nevertheless, the core will generally be quite hot and the solid layer very thin.  We neglect any solid phases in our calculations below.  The gas states are well described by the ideal gas equation of state
\beqa
P = \frac{\rho\,k_B T}{M_{\rm gas}} \,,
\eeqa
where $M_{\rm gas}$ is the mass of the gas particles, either $m_p$ for atomic gas or $2 m_p$ for molecular gas. Here, $m_p$ is the proton mass. For the gas far away from the axion star, it has an ideal gas equation of state with the temperature given by $T_m$. At early times, the interactions of free electrons with CMB photons via Compton scattering maintains a uniform gas temperature that follows the CMB photon temperature. At later times, after redshift $z_t\approx 137$~\cite{Barkana:2000fd}, corresponding to a CMB photon temperature $T^\gamma_t \approx 0.032~\mbox{eV}$, the gas temperature declines adiabatically as $T_m = [(1+z)/(1+z_t)]^2\,T^\gamma_t$ and is colder than CMB photons until  photoionization heating when the first stars start to burn. 

For the liquid states surrounding the axion star, we assume a common-power law equation of state
\beqa
\label{eq:eos-interior}
P = \frac{\rho\, k_B T}{2 \,m_p} \,\left(\frac{\rho}{\rho_0}\right)^{\gamma_{\rho}} \,\left(\frac{T}{T_0}\right)^{\gamma_{_T}} \,.
\eeqa
We fit the parameters to a combination of calculations and data~\cite{NIST,Morales20072010}, finding
\beqa
\rho_0 = 51.33~{\rm kg }/{\rm m}^3, \quad T_0 = 1000~{\rm K}, \quad \gamma_{\rho} = 1.36, \quad \gamma_{_T} = -0.89 \,.
\eeqa
We approximate the gas to liquid phase transition as a first order transition at $(\rho/\rho_0)^{\gamma_{\rho}} (T/T_0)^{\gamma_{_T}} = 1$.  This transition essentially defines the boundary of the baryon component of the axion star, as the density of the baryons grows very rapidly near the boundary, generally surpassing the required density to reach a liquid metallic state.  We therefore define $R_{\rm bound}$ to be the radius for which 
\beqa
\label{eq:dens-boundary}
\left[\frac{\rho(R_{\rm bound})}{\rho_0}\right]^{\gamma_\rho} \left(\frac{T_{\rm surf}}{T_0}\right)^{\gamma_{_T}} = 1 \,,
\eeqa
where $T_{\rm surf}$ is the surface temperature of the axion star, which we revisit below.

The metallic liquid phase has a large DC conductivity of order $\sigma_0 \sim 2500~{\rm S} / {\rm cm}$~\cite{PhysRevLett.116.255501} (in natural units, ${\rm S} / {\rm cm} = 0.0075$~eV). The conductivity can be related to the absorption cross-section for axions.  The details of this calculation are presented in the Appendix~\ref{appendix} and have been studied by Refs.~\cite{Pospelov:2008jk,Hochberg:2015fth,Hochberg:2016ajh}.  If the axions are sufficiently light, then frequency dependent effects in the conductivity can be neglected.  The relaxation rate in metallic hydrogen is quite large, such that the onset of mass dependence should be well above our benchmark masses \cite{2015arXiv150400271V}.  We make this approximation below.  For the two axion benchmarks, this leads to power generated in the core of order (see Appendix~\ref{appendix} for the relation of conductivity $\sigma_0$ and the axion absorption rate $\Gamma^a_{\rm abs.}$)
\beqa
\renewcommand{\arraystretch}{1.5}
P_{a\odot} = M_{a\odot} c^2 \,\Gamma^a_{\rm abs.} = M_{a\odot} c^2 \, \frac{3\,m_a^2}{4\,\pi\,\alpha \,f_a^2}\, \frac{\sigma_0}{\epsilon_0} \approx 
\bigg\{
\begin{array}{ll}
 2\times 10^{13}~{\rm W}  &\quad  \mbox{for}\;\; m_a  =5~\mbox{meV} \,,  \\
  2\times 10^{5}~{\rm W}   &\quad   \mbox{for}\;\; m_a = 0.05~\mbox{meV} \,.
\end{array}
\eeqa
The generated power is emitted as blackbody radiation near the surface of the HAS. Heat transportation within the HAS is rather complex and depends on the exact phase of the matter inside the HAS as well as the heat transportation properties. It is, however, likely that radiative or convective transportation will dominate. 
We do not need to solve this exactly, but rather can approximate the HAS as a black body, solving simply for the boundary condition that requires that the total generated power is equal to the total power emitted at the surface
\beqa
\label{eq:temp-boundary}
P_{a\odot} = 4 \,\pi \,R_{\rm bound}^2 \,\epsilon \,\sigma_{_{\rm SB}}  \,T_{\rm surf}^4 \,,
\eeqa
where $T_{\rm surf}$ is the surface temperature of the HAS; $\sigma_{_{\rm SB}}$ is the Stefan-Boltzmann constant; the emissivity $\epsilon$ is chosen to be  $\epsilon=1$. In order to do so, we must assume that the equations of state inside the HAS remain unchanged. If the temperature inside the HAS were to get so high that the fluid inside undergoes an additional phase transition, say to a plasma, then the total power generated could be modified. 

We leave the detailed properties of the interior of the HAS, such as the exact temperature profile and the opacity, undetermined, assuming only that we know the density profile and conductivity.  The conductivity is assumed to be that of fluid metallic hydrogen as determined by Ref.~\cite{PhysRevLett.116.255501}.  We also assume the fluid hydrogen equation of state Eq.~\eqref{eq:eos-interior} holds throughout the interior of the HAS.  Aside from ensuring that our approximations regarding the density and conductivity are justified, the structure of the interior of the HAS has little bearing on the main observables we consider.

\begin{figure}[!tb]
\centering
\includegraphics[width=0.6\textwidth]{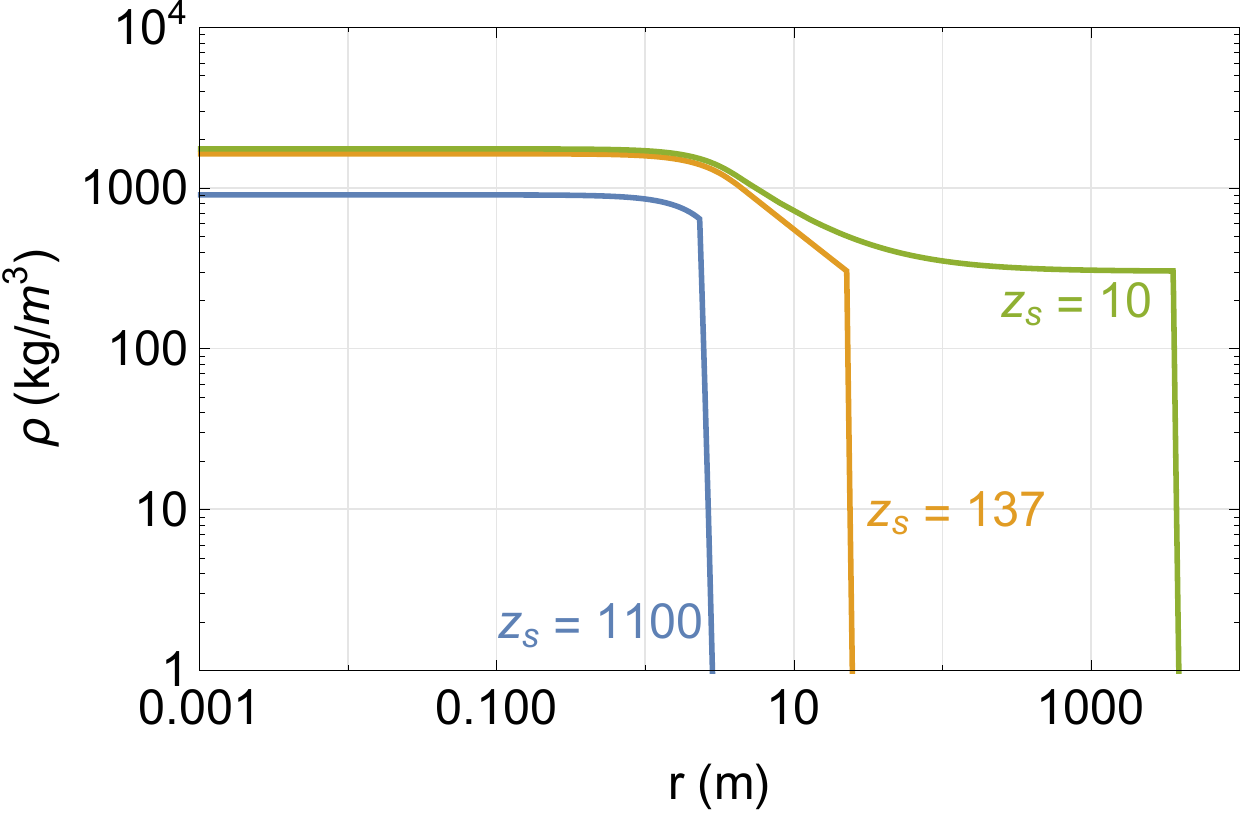}
\caption{Density profile of the fluid cloud around an axion star at several possible final redshifts. At $z_s \approx 137$, the gas around the HAS thermally decouples from the photon background.  At $z_s \approx 10$, galactic and stellar structure forms, which may disrupt the accretion process.  The axion star mass is $M_{a\odot} = 3.6 \times 10^{19}~{\rm kg}= 1.8\times10^{-11}M_{\odot}$ and it has a radius of around 5 m.  The baryonic cloud is a subdominant contribution to the total mass of the HAS, increasing from $3.9 \times 10^4~{\rm kg}$ for $z_s = 1100$ to $1.2 \times 10^{13}~{\rm kg}$ for $z_s = 10$.}\label{fig:density}
\end{figure}

Given our condition for the boundary of the HAS Eq.~\eqref{eq:dens-boundary} and Eq.~\eqref{eq:temp-boundary}, we solve for both the boundary radius $R_{\rm bound}$ and the surface temperature $T_{\rm surf}$, as well as the density profile of the fluid in the HAS.  From our analysis above, it is not clear when to stop evolving the HAS in the expanding universe.  As such, we define a redshift $z_s$ at which the evolution stops and present several different possibilities in our solutions, corresponding to the times of recombination at $z = 1100$, matter-photon kinetic decoupling at $z \approx 137$, and reionization and star formation at $z \sim 10$. Our solution for the profile of hydrogen cloud is shown for different values of $z_s$ in Fig.~\ref{fig:density} for an axion mass of $m_a c^2 = 5 \times 10^{-3}~{\rm eV}$.  A summary of the most important properties of the cloud is shown Table \ref{tab:boundary} for two different benchmark QCD axion masses.
\begin{table}[!ht]
\renewcommand{\arraystretch}{1.5}
\centering
\begin{tabular}{c | c c || c c}
\hline\hline
 $m_a$   &    \multicolumn{2}{c||}{$5~\mbox{meV}$}   & \multicolumn{2}{c}{$0.05~\mbox{meV}$} \\   \hline
$z_s$ &  $R_{\rm bound}$ (m) & $T_{\rm surf}$ (K) & $R_{\rm bound}$ (m) & $T_{\rm surf}$ (K)\\
\hline
1100 &   2.3 & $4.7 \times 10^4$  &   2.6 & 450 \\
137 & 23 & $1.5 \times 10^4$ & 24 & 150 \\
10 &  3700 & 1200  &  3800 & 12 \\
\hline\hline
\end{tabular}
\caption{Properties of the baryonic cloud boundary surrounding an axion star at different stopping redshifts in the evolution of the universe. }\label{tab:boundary}
\end{table}

At times later than roughly redshift $z = 10$, the surroundings of different HAS become vastly different.  The properties of the HAS may then depend on where the axion is located.  In particular, effects like collapse into galaxies and reionization could have a significant effect on the clouds surrounding axion stars.  We therefore do not attempt to evolve the HAS beyond this point and ascribe a substantial uncertainty to the final surface temperature of the cloud.  It is also worth noting that if the HAS were to collide with a molecular cloud, it could be significantly disturbed in a way reminiscent of the collision of bullet clusters.  The HAS cloud would likely get absorbed into the molecular cloud, while the axion core would move through unperturbed.  To see this, we perform a rough estimate of the scales involved here.  For a HAS near the Earth, the velocity relative to any nearby objects is of order $100~{\rm km}/{\rm s}$, so that the kinetic energy transfer in a collision with a gas molecule is of order $\Delta E \sim m_{\rm H} v^2 \sim 2 \times10^{-17}~{\rm J}$.  Meanwhile, the binding energy for a gas molecule to the axion star is of order $E_b \lesssim G_N M_{a\odot} m_H / R_{a\odot}$, that is $E_{\rm b} \lesssim 8 \times 10^{-19}~{\rm J}$, which is much less.  It remains to determine the probability that a collision between a HAS and an unrelated gas cloud leads to most of the hydrogen in the HAS interacting.  The number of interactions for a single hydrogen atom of HAS to interact with the gas cloud is of order $\pi\,a_0^2\, R_{_{\rm GC}} n_{_{\rm GC}}$, where $a_0$ is the hydrogen Bohr radius, $R_{_{\rm GC}}$ is the radius of the gas cloud, $n_{\rm GC}$ is the density of the gas cloud.  To get an idea of the probability of destruction in a given event, we consider two benchmark cases.  First, consider a collision with a giant molecular cloud, which has a density of $10^4$--$10^6$~${\rm cm}^{-3}$ and a radius of 2--100 parsecs~\cite{2011ApJ...729..133M}.  The resulting number of interactions per hydrogen atom is then at least roughly $5 \times 10^6$, so that all the hydrogen in the HAS is likely to interact and disrupt the hydrogen cloud of the axion star.  For a smaller system, like a solar system which has a diffuse density of order $5~{\rm cm}^{-3}$ and a radius of order 100 AU~\cite{Nieto:2003qy}, the expected number of interactions is order 1, meaning that an appreciable fraction of the baryonic cloud may survive such an interaction. Therefore, a fly-by HAS is still possible to behave as a point-like source.

\section{HAS Detection}
\label{sec:detection}
Radiation from the hot hydrogen fluid surrounding axion stars could be seen by a suitably high resolution telescope.  Provided that the HAS fluid reaches a sufficient density to become metallic and that the metallic fluid reaches the edge of the axion core, the power generated by the HAS is independent of the radius of the baryonic cloud.  In particular this means that the total power emitted is insensitive to the ultimate surface temperature of the HAS.  On the other hand, the hotter the HAS, the lower wavelength (higher frequency) the emitted radiation will be.  Depending on the final temperature of the HAS, the ideal wavelength for a sensitive telescope is in the UV through IR range. 

\begin{figure}[!htb]
\centering
\includegraphics[width=0.45\textwidth]{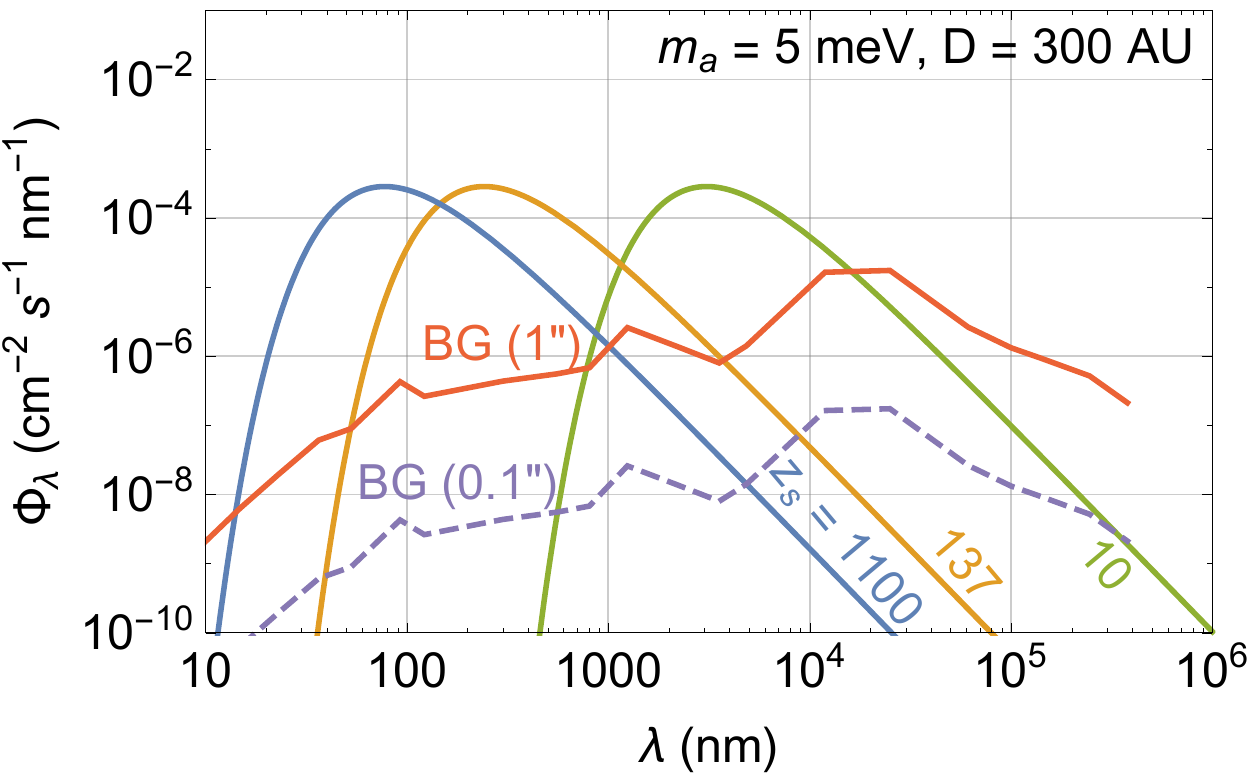} \hspace{3mm}
\includegraphics[width=0.45\textwidth]{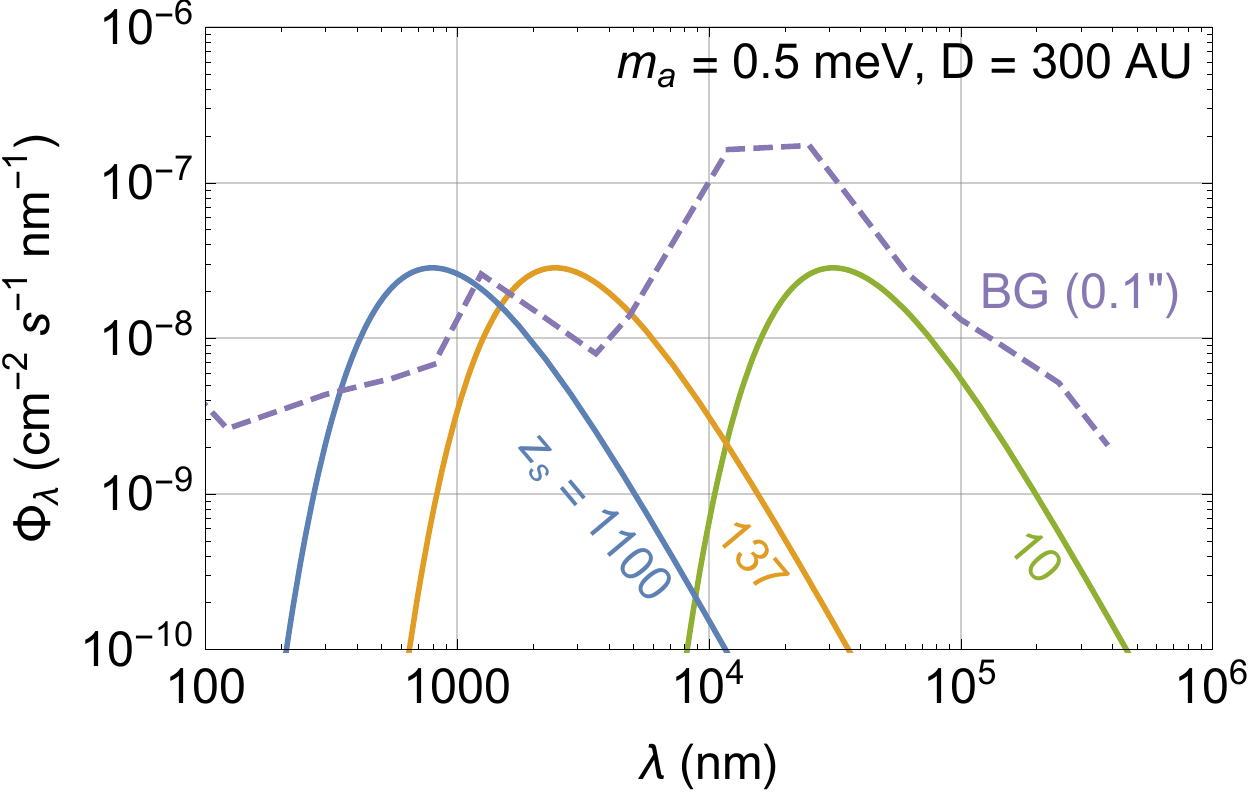}
\caption{Flux of photons from a nearby HAS for several different evolving-stop redshifts, as well as the diffuse background flux for different assumptions about telescope angular resolution~\cite{Henry:1999zw}. We present the signal flux for two different choices of axion mass, $m_a = 5~{\rm meV}$ (left panel) and $m_a = 0.5~{\rm meV}$ (right panel).  The total flux scales as $m_a^4$ for a fixed $z_s$, while the temperature for a lower axion mass is reduced due to decreased power generation.  As a result, the peak wavelength is larger for a lower mass.
\label{fig:flux}}
\end{figure}

Given an HAS at a distance $D$ from the detector, surface temperature $T$, and baryonic cloud radius $R_{\rm bound}$, the rate of photons passing through a detector is given by
\beqa
\Phi_\lambda= \frac{d\Phi}{d\lambda} \equiv \frac{dN_\gamma}{dt\,dA\,d\lambda} = \left(\frac{R_{\rm bound}}{D}\right)^2 \frac{2\,\pi c}{\lambda^4} \frac{1}{e^{h\,c / \lambda\, k_B\,T_{\rm surf}} -1} \,,
\eeqa
There are several different strategies one could use to search for HAS photon emissions.  One could look for a nearby HAS, which should appear as a fairly bright point source in the sky.  The {\it apparent magnitude} (defined with respect to the brightness of Vega)  of such a source would be
\beqa
{\mathbf m} &=& 18.9 - 2.5 \log_{10} \left[\left(\frac{P}{2 \times 10^{13}~\mbox{W}}\right) \left(\frac{300\,\mbox{AU}}{D}\right)^2\right]  \nonumber \\
&=& 
33.0 - 2.5 \log_{10} \left[\left(\frac{P}{2 \times 10^{13}~\mbox{W}}\right) \left(\frac{1\,\mbox{pc}}{D}\right)^2\right] 
\,.
\eeqa
Given the local density of dark matter, we expect there to be axion stars and HAS within roughly $300~{\rm AU}$ of the Earth, which is near enough to give a detectable flux.  If the telescope points away from the galactic plane, the estimated diffuse background can be relatively small~\cite{Henry:1999zw}.  The flux for a nearby HAS is shown in Fig.~\ref{fig:flux}. Depending on the telescope angular resolution and observation time, the HAS located at a distance of the solar-system size can be observed. 

One could also look at an area of the sky rich in dark matter and low in background.  The best such option given the frequencies emitted by the HAS is a baryon-poor dwarf spheroidal galaxy.  The dwarf spheroidal with the smallest luminosity to mass ratio is the Ursa Major II dwarf, which has a luminosity of  $\sim4000\,L_\odot$, a distance of $30~{\rm kpc}$ and a mass of around $5 \times 10^6 M_\odot$~\cite{Simon:2007dq,Martin:2008wj}.  Given that mass, there are a total of roughly $2.8 \times 10^{17}$ axion stars in the galaxy, leading to a total luminosity of 
\beqa
L_{a\odot}^{\rm UMaII} \sim 1.4\times10^4 \, L_\odot \,\times\,\left(\frac{1.8 \times 10^{-11} M_\odot}{M_{a\odot}}\right)\left( \frac{P_{a\odot}}{2 \times 10^{13}~{\rm W}}\right)\,.
\eeqa
For our benchmark model, this luminosity is comparable to that of the dwarf galaxy, so a study of the spectrum of dwarf spheroidal could be sensitive to HAS.

Finally, one could simply look for diffuse light coming from HAS.  The constraints from searches for diffuse photons should be fairly weak.  To see that this is the case, we perform a simplified conversion of HAS radiation to effective dark matter decay into a pair of photons.  While a dark matter particle decaying to a pair of photons would yield monochromatic photons of energy $m_{\rm DM} / 2$, our axion decays to photons with energy peaked at a temperature $T \sim 1000~{\rm K} \,\mbox{--}\, 50,000~{\rm K}$ corresponding to photons of energy roughly $k_B T \sim 0.09~{\rm eV} \,\mbox{--}\, 4.3~{\rm eV}$, such that we can take the effective mass to be $m_{\rm eff} c^2 \sim 0.18~{\rm eV} \,\mbox{--}\, 8.6~{\rm eV}$.  The total rate of photon emission per axion is given by roughly the total power emitted divided by the typical energy of emitted photons and the total number of axions in the axion star, $\Gamma_{\rm eff} \simeq P_{a\odot} / (k_B \,T \,\mathcal{N}_a)$, giving an effective lifetime in the range $\tau_{\rm eff} \sim 2.9 \times 10^{24}~{\rm s} \,\mbox{--}\, 1.5 \times 10^{26}~{\rm s}$.  Constraints in the $m_a$--$\tau_a$ plane for monochromatic photons have been presented in Ref.~\cite{Arias:2012az}.  No constraint is found for the effective mass and lifetime ranges above.

\section{Axion Stars in Planets}
\label{sec:axion-planets}
Axion stars are have fairly low masses and, as such, would be small gravitational sources in a mature solar system.  A remote axion star, isolated from baryons would be a small gravitational source and would only be potentially observable by decay into photons \cite{Eby:2015hyx,Eby:2016cnq,Braaten:2016dlp}.  In this work, we have focused on signals from axion stars surrounded by baryonic matter. We have already discussed the potential for observing the hot baryonic cloud around a nearby axion star in the previous section, but another possibility is that axion stars end up inside planets or stars themselves.  This scenario is less likely, as the processes for axion star capture are fairly weak and a planetary collision with other objects could cause the axion star to be pushed away from the center of the planet.  In this section, we nevertheless briefly discuss the prospects and signals for such a scenario.

An axion star that ended up in a proto-stellar cloud would likely get stripped of its baryonic cloud.  It therefore would have few means to lose energy and end up in the center of a planet.  There are processes with three body gravitational interactions that could cause a (proto-)solar system to capture an axion star \cite{1991ApJ...368..610G,Lundberg:2004dn,Peter:2009mm,Peter:2009qj,2009IJMPD..18.1903K,Khriplovich:2010hn,2010arXiv1004.5258E}.  With some degree of uncertainty, the conclusion of studies of such processes indicate a fairly small abundance of bound dark matter.  If the axion star did end up captured by the proto-stellar cloud, it could end up seeding the formation of a planet, or the star itself, as it would be a large density perturbation within the cloud~\cite{Armitage:2007gn}.  As seen above and in Appendix \ref{appendix}, axions can convert off conduction band electrons into phonons and generate heat.  A planet containing an axion star would have an additional heat source due to axion conversion in the metallic planetary core, which is likely to be a conductor.  The heating occurs much as in a HAS, except with a smaller conductivity for which the frequency dependence becomes important. 

\begin{figure}[ht!b]
  \centering
\includegraphics[width=0.5\textwidth]{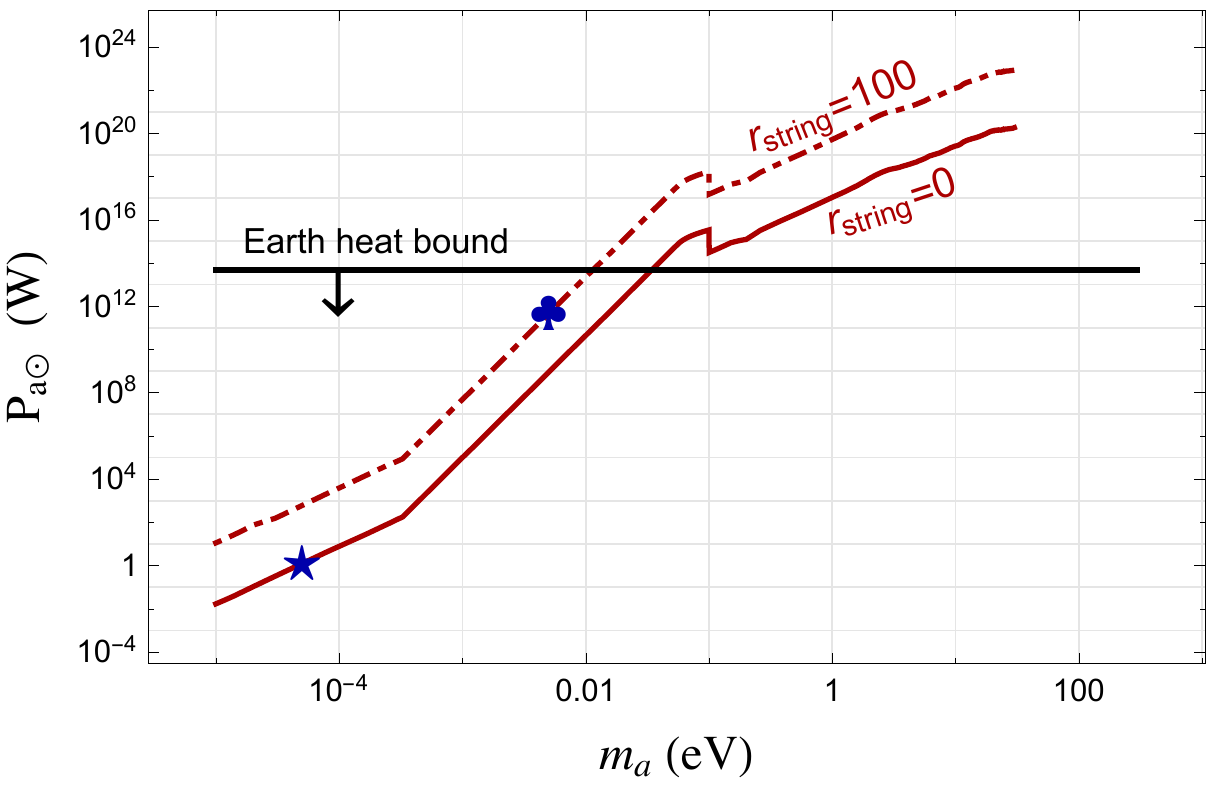} 
  \caption{Power generated by an axion star converting in the core of the Earth compared to the internal heat $47\pm2$~TW in the Earth~\cite{se-1-5-2010}.}
  \label{fig:sigmavn-axion}
\end{figure}

If an axion star was at the core of the Earth, the heat generated by axion conversions could be comparable to the internal heat, $47\pm2$~TW, generated by radioactive decays and primordial heat in the Earth~\cite{se-1-5-2010}.  This is particularly true for high mass axions, which have an enhanced conversion rate in iron.  The relaxation rate in iron is significantly lower than in hydrogen such that frequency dependence is important for even the QCD axion.  We therefore use the frequency-dependent conductivity for iron~\cite{haynes2015crc} to estimate the power generated by an axion star in the core of the Earth, which is shown in Fig.~\ref{fig:sigmavn-axion}. For a heavier axion mass of 5~meV, the additional heat from axion absorption is around 1~TW and just one order of magnitude below the Earth total internal heat. For a lighter axion mass of 0.05~meV, both the suppression factor of $m_a^2/f_a^2$ and the frequency-dependent Iron conductivity make the generated heat negligible. If the axion stars exist in other smaller size planets, exoplanets or moons, the axion-star-generated heat could be easier to notice. 

In addition, the presence of an axion star at the center of a planet would alter the density-radius relation probed by seismology.  By measuring seismic wave propagation at various points along the surface of the Earth, a reasonably accurate picture of the interior of the Earth has been inferred \cite{DZIEWONSKI1981297,Kennett01071995}.  With sufficient resolution of the core, a large enough discrepancy between observation and reasonable models could be detected.  Such a discrepancy could be explained by an axion star core, which would tend to increase the density of baryonic matter in the core of the axion star.

\section{Discussion}
\label{sec:discussion}

We have found a very interesting and new class of astrophysical objects based on the well-motivated QCD axion model, but there remain several largely uncertain components of our analysis.  The preferred axion parameter space depends both on a better understanding of the thermal axion mass and the generation of axions from the decay of topological defects.  The choices of axion mass and decay constant have large, fourth power effect on the power generated by axion star interactions with a surrounding metal, so the viability of the signal presented in this work relies on a large axion mass implied by copious production in axion string decays.  In addition, a complete understanding of the formation of primordial axion stars requires a full numerical simulation of axions around the time of the QCD phase transition.  We have assumed that all the axions collapse into axion stars for our calculations, but axion star formation efficiency will be less than one.  Some simulations have shown a formation efficiency of 30\% to 50\% \cite{Khlebnikov:1999qy}, but it would be interesting to verify this finding with modern computing power and techniques.  Furthermore, the process of collapse could cause the axion star to shed axions or to collapse to an even more dense state than that assumed in this work.  Other possibilities due to axion self-interactions, such as a bosenova \cite{Eby:2016cnq}, also cannot be ruled out at this time.  Finally, though the early evolution of the HAS is relatively easy to understand using quasi-static evolution of a system in hydrostatic and radiative equilibrium, the evolution of HAS beyond the time of galaxy and star formation remains challenging.  Reionization may heat up the gas cloud surrounding the axion star.  Presence of gas with heavier atoms and molecules surrounding the axion star could cause it to accrete more gas or other massive particles.  The reverse is also possible: a surrounding hot gas could cause the axion star to partially evaporate.  In addition, the axion star is fairly fragile to collisions with an unrelated dense cloud of gas, which could strip the baryonic axion star cloud as the axion star core passes through undisturbed.  A complete study of such astrophysical effects is beyond the scope of this work.

Despite the large number of open questions, we believe that signals due to hot matter surrounding axion stars deserve exploration using telescopes.  We have found that axion stars could be quite nearby, within a few hundred astronomic units of the Earth.  This leads to a point source signal that should be visible with high resolution telescopes.  Due to the large uncertainty in the temperature of the axion stars, a wide frequency coverage is in order.  We have found that the blackbody peak frequency may be in the range of ultraviolet to infrared if the baryonic cloud properties are frozen in at some early time.  A ``smoking gun'' for ascertaining our signal would be HAS emissions from a low-baryon region of the sky, such as a dwarf spheroidal galaxy.  Under certain assumptions, the total expected axion star luminosity in a dwarf galaxy could be comparable to or larger than the total observed luminosity, leading to constraints on the axion star properties.

\section{Conclusions}
\label{sec:conclusions}

The QCD axion remains a compelling solution to both the strong CP and dark matter puzzles. We have found that on small distance scales, the cosmology of the axion is interesting and leads to new potential signals. In particular, the following picture of QCD axion cosmology emerges:
\begin{itemize}
\item Before the QCD phase transition, the axion field is uniform on roughly the Hubble scale, with a network of axion strings that are decaying into axions.

\item Just before the QCD phase transition, the thermal axion mass becomes important and causes both the axion field to roll down to its minimum, producing axions, and the remaining long axion strings to annihilate away.

\item After the QCD phase transition, the comoving density of axions is frozen, but there are large density contrasts on the scale of Hubble, around 10 km.

\item The axions can rapidly relax down to their local ground state, forming a compact, dense, meter scale BEC, called axion stars.  The typical mass of the axion star is roughly $10^{-11}\,M_{\odot} \approx 10^{19}~{\rm kg}$.

\item Around the time of recombination, the primordial hydrogen gas begins to collapse into the gravitational potential well of the axion stars, reaching a dense, metallic fluid state and forming combined objects called Hydrogen Axion Stars (HAS).

\item In this epoch, axions can interact with conduction band electrons, generating heat in the hydrogen fluid and causing it to blackbody radiate with a peak wavelength in the UV.

\item As the universe expands, the gas surrounding the axion stars cools and further accretes onto the HAS, causing the hydrogen fluid to expand and cool.  The peak wavelength of the radiation increases toward the IR.  The mass of the hydrogen cloud is subdominant to that of the axions in the HAS, but can grow to be of order $10^{13}~{\rm kg}$, comparable to the mass of an asteroid, but distinguishable by its composition.  The radius can grow to be several kilometers.

\item At late times, when reionization and star formation occur, the properties of the HAS depend on its local surroundings and the fate of the HAS becomes more uncertain.  The final state of the HAS depends on the redshift at which the HAS evolution stops.
\end{itemize}
The HAS is likely to be fairly hot and to have sufficient power output to be observed by telescopes sensitive to photons in the UV to IR range, yielding observational signals.  In our local vicinity, axion stars making up a predominant component of the dark matter would have of a number density of order $(300~{\rm AU})^{-3}$, such that there can be nearby HAS visible as point sources to high resolution telescopes.  In addition, the combined HAS luminosity from faint dwarf spheroidal galaxies such as Ursa Major II (with order $10^{17}$ HAS) can be comparable to the observed luminosity of such objects.  A rich new phenomenology is open for exploration and discovery.

\subsection*{Acknowledgments}
We would like to thank Eric Braaten, Dan Chung, Madelyn Leembruggen, Peter Suranyi, Rohana Wijewardhana, and Yue Zhao for discussion and comments. This work is supported by the U. S. Department of Energy under the contract DE-FG-02-95ER40896. This work was partially performed at the Aspen Center for Physics, which is supported by National Science Foundation grant PHY-1066293. We could like to acknowledge the Mainz Institute for Theoretical Physics (MITP) for enabling us to complete a portion of this work. 

\appendix
\section{Absorption of Axions in Metals}
\label{appendix}

In this appendix, we determine the absorption cross-section for photons
in a material and relate this to the absorption cross-section for an
axion in the same material.

The absorption cross-section of photons in the material is related to
the macroscopic electromagnetic properties of the material, which can
be parametrized in terms of the complex dielectric constant, index of
refraction, or conductivity (see Ref.~\cite{schrieffer1983theory,dressel2002electrodynamics,An:2013yua} for detail).  We choose to work with the latter for
reasons that will become clear soon.

We need to obtain a relation between the conductivity $\hat{\sigma} =
\sigma_1 + i \sigma_2$ and the absorption cross-section $\sigma_{\rm
  abs}$.  The approach we use is the optical theorem.  For a particle $P$
with four-momentum $q$ (not necessarily on-shell), the optical theorem
says that
\begin{equation}
  \label{eq:1}
  {\rm Im}\,\mathcal{M}[P(q) \to P(q)]= \sqrt{q^2}\, \Gamma(q) \,,
\end{equation}
where $\Gamma$ is the decay rate of $P$.  The amplitude
$\mathcal{M}[P(q) \to P(q)]$ equals to the product of the polarization tensor
$\Pi^{\mu\nu}(q) = e^2 \int d^4x\, e^{i q \cdot x}
\langle 0| T[J^{\mu\dagger}(x) J^\nu(0)] |0\rangle$ and the
corresponding polarization vectors for the state.~\footnote{This
  relation can be seen by explicitly constructing the leading-order
  $1 \to 1$ amplitude, which arises at second order in perturbation
  theory ({\it i.e.} it goes like the $H_I^2 \supset J^2$).}  We can
therefore write the polarization tensor as
\begin{equation}
  \label{eq:2}
  \Pi^{\mu\nu}(q) = \epsilon_L^{\mu *} \epsilon_L^{\nu}\, \Pi_L(q) + \sum_T
  \epsilon_T^{\mu *} \epsilon_T^{\nu} \, \Pi_T(q)   \,,
\end{equation}
where $\epsilon_{L,T}$ are longitudinal and transverse polarization
four vectors respectively, satisfying $q \cdot \epsilon_{L,T} = 0$.
Thus, if we determine the polarization tensor for the photon in the
medium, we can determine the rate $\Gamma$ at which photons get
absorbed by the medium.

To determine the polarization tensor, we begin by determining a
similar quantity, $R^{\mu\nu}(x,y)$, the kernel that transforms $A$
into $J$
\begin{equation}
  \label{eq:3}
  -e \langle J^\mu(x) \rangle = - \int d^4y R^{\mu\nu}(x,y) \langle
  A_\nu(y) \rangle   \,.
\end{equation}
In momentum space, this relation can be written as
\begin{equation}
  \label{eq:4}
  e J^\mu(q) = R^{\mu\nu}(q) A_\nu(q)  \,,
\end{equation}
since $R^{\mu\nu}(x,y)$ can be written as a function of $x - y$.  The kernel can be determined by explicitly
calculating the expectation value of the current operator $J$ as
\begin{equation}
  \label{eq:5}
  \langle \Omega(t) | J^\mu(x) | \Omega(t) \rangle = \langle 0 |
  U(\infty,t) J^\mu(x) U(t,-\infty) |0\rangle  \,,
\end{equation}
where $|\Omega(t) \rangle$ is the time-evolved vacuum state,
$|0\rangle$ is the vacuum state at $t \to \pm \infty$, and $U$ is the
usual interaction picture evolution operator
\begin{equation}
  \label{eq:6}
  U(t,t') = \mbox{T} \exp\left[-i \int_{t^\prime}^t H_{\rm
      int}(t^{\prime\prime})  dt^{\prime\prime}\right] = 1 - i
  \int_{t^\prime}^t H_{\rm int}(t^{\prime\prime} dt^{\prime\prime}) +
  \dots\, .
\end{equation}
Since $U$ is unitary and the vacuum state at $t \to \pm \infty$ is the
same, we can write $U(\infty,t) = U^\dagger(t, -\infty)$, so that at
leading order in perturbation theory we find
\begin{equation}
  \label{eq:7}
  \langle J^\mu \rangle = -i \langle 0 | \left[J_\mu(x),
    \int_{-\infty}^t H_{\rm int}(t^\prime) dt^\prime\right] | 0 \rangle  \,.
\end{equation}
The interaction Lagrangian is $H_{\rm int} = -e \int d^3x  J^\nu(x)
A_\nu(x)$ as usual and we can write $\int_{-\infty}^t dt^\prime = \int
dt^\prime \theta(t - t^\prime)$, so that
\begin{equation}
  \label{eq:8}
  R^{\mu\nu} = - i \,e^2\, \langle 0 | [J^\mu(x), J^\nu(y)] |0\rangle
  \theta(x^0 - y^0) \,.
\end{equation}
We can see that this is closely related to the polarization tensor,
differing just by the operator ordering.  In momentum space, this
corresponds to a difference in the prescription for dealing with the
poles, such that
\begin{equation}
  \label{eq:9}
  {\rm Re} \,R^{\mu\nu}(q) = {\rm Re} \,\Pi^{\mu\nu}(q) \,, \qquad  {\rm
    Im}\,R^{\mu\nu}(q) = {\rm sgn}(q^0)\, {\rm Im}\,\Pi^{\mu\nu}(q) \,.
\end{equation}
Since we are only interested in $q^0 > 0$ here, we can just determine
$R^{\mu\nu}$.

To do so, we first rewrite $J^\mu$ and $A^\mu$ making use of charge
conservation and gauge invariance respectively.  First, in momentum
space, the continuity equation allows us to write
\begin{equation}
  \label{eq:10}
  J^0 = \frac{\mathbf{q} \cdot \mathbf{J}}{q^0} \,.
\end{equation}
The current density $\mathbf{J}$, by the definition of conductivity,
is given by $-e \mathbf{J} = \hat{\sigma} \mathbf{E}$, where
$\hat{\sigma} = \sigma_1 + i \sigma_2$.  Furthermore, the electric
field $\mathbf{E} = - i (\mathbf{q}\,A^0 - q^0\, \mathbf{A})$, so that
the current four-vector is
\begin{equation}
  \label{eq:11}
 -e  J^\mu = -i\,\hat{\sigma} \left(\frac{\mathbf{q}^2}{q^0} A^0 - \mathbf{q} \cdot
    \mathbf{A}, \mathbf{q} \,A^0 - q^0 \,\mathbf{A}\right).
\end{equation}
Since we are calculating a physical quantity ultimately, our final
result should be gauge invariant.  Nevertheless, it is helpful to pick
a gauge for our calculation.  One convenient gauge is the Coulomb gauge
with $\mathbf{q} \cdot \mathbf{A} = 0$, so that $-e J^0 = -i
\hat{\sigma} \mathbf{q}^2
A^0 /q^0$, $\boldsymbol{\epsilon}_L \cdot \mathbf{A} = 0$ and $\epsilon_L
\cdot A = \epsilon_L^{0} A^0 = |\mathbf{q}| A^0 / \sqrt{q^2}$.  We can
then go ahead and solve the various components of Eq.~\eqref{eq:4}.  We
begin with the 0 component.  Since $\epsilon_T^{0} = 0$, we have
\begin{equation}
  \label{eq:12}
  -i \hat{\sigma} \frac{\mathbf{q}^2}{q^0} A^0 = - \Pi_L \frac{\mathbf{q}^2}{q^2} A^0 \,,
\end{equation}
leading to
\begin{equation}
  \label{eq:13}
  \Pi_L = i \hat{\sigma} \frac{q^2}{q^0} \,.
\end{equation}
For the spatial components, with our choice of gauge, 
\begin{equation}
  \label{eq:14}
  i\hat{\sigma}( \mathbf{q} A^0 - q^0 \mathbf{A}) = \Pi_L \boldsymbol{\epsilon}_L
  \epsilon_L^{0} A^0 - \Pi_T \sum_T \boldsymbol{\epsilon}_T
  (\boldsymbol{\epsilon}_T \cdot \mathbf{A}) \,.
\end{equation}
Since we have
\begin{equation}
  \label{eq:15}
  \sum_\lambda \boldsymbol{\epsilon}_\lambda \boldsymbol{\epsilon}_\lambda = \boldsymbol{\epsilon}_L
  \boldsymbol{\epsilon}_L + \sum_T \boldsymbol{\epsilon}_T
  \boldsymbol{\epsilon}_T = 1 + \frac{\mathbf{q}\,\mathbf{q}}{q^2} \,,
\end{equation}
as usual for a vector polarization and $\mathbf{q} \cdot \mathbf{A} =
0$, we find
\begin{equation}
  \label{eq:16}
  i\hat{\sigma}( \mathbf{q} A^0 - q^0 \mathbf{A}) =  i
  \hat{\sigma} \mathbf{q} A^0 - \Pi_T \mathbf{A}   \,,
\end{equation}
leading to
\begin{equation}
  \label{eq:17}
  \Pi_T = i \hat{\sigma} q^0.
\end{equation}

Assuming that $\mathbf{q}$ is real, we find the decay rates from
Eq.~\eqref{eq:1} for longitudinal and transverse modes are given by
\begin{equation}
  \label{eq:18}
  \Gamma_L = \sigma_1 \frac{\sqrt{q^2}}{q^0} \,, \qquad \Gamma_T =
  \sigma_1 \frac{q^0}{\sqrt{q^2}} \,.
\end{equation}
Further, if we take the approximation that $|\mathbf{q}| \ll q^0$,
which is certainly true for the kinematics of axion absorption, we
find that
\begin{equation}
  \label{eq:19}
  \Gamma_{{\rm abs.}}^\gamma = \Gamma_L = \Gamma_T = \sigma_1  \,.
\end{equation}
In terms of the absorption cross-section in the medium, which we
assume is coming entirely from absorption by free electrons in the
conductor, we find that
\begin{equation}
  \label{eq:20}
  \Gamma_{{\rm abs.}}^\gamma = \langle n_e \sigma_{\rm abs} v_{\rm rel}\rangle \,,
\end{equation}
where $n_e$ is the number density of free electrons, $\sigma_{\rm
  abs}$ is the absorption cross-section, and $v_{\rm rel}$ is the
relative velocity of the incoming particles.

Given the photon absorption rate in the material, the axion absorption rate is determined by a simple rescaling \cite{Pospelov:2008jk}
\beqa
\label{eq:axion-absorption}
\Gamma_{{\rm abs.}}^a = \frac{3\,m_a^2}{4 \pi \alpha f_a^2} \Gamma_{{\rm abs.}}^\gamma \,,
\eeqa
where $\alpha$ is the fine-structure constant.  Eq.~\eqref{eq:19} thus suffices to determine the axion absorption rate in a metal once the conductivity is known.

\bibliography{axion}
\bibliographystyle{JHEP}

\end{document}